\documentclass[8pt,a4paper,twocolumn]{article}
\usepackage[margin=20mm]{geometry}
\usepackage{amsmath,amssymb,amsfonts}
\usepackage{algorithmic}
\usepackage{graphicx}
\usepackage{textcomp}
\usepackage{multirow}
\usepackage{tabularx}
\usepackage{cases}
\usepackage{empheq}
\usepackage{algorithm}
\usepackage{listings}
\usepackage{wrapfig}
\usepackage{url}
\usepackage{here}

\begin{document}

\title{Efficient and Scalable Architecture for Multiple-chip Implementation of Simulated Bifurcation Machines}

\author{Tomoya Kashimata, Masaya Yamasaki, Ryo Hidaka, and Kosuke Tatsumura$^{\ast}$\\
\small Corporate Research and Development Center, Toshiba Corporation, \\
\small  1 Komukai Toshiba-cho, Saiwai-ku, Kawasaki 212-8582, Japan\\
\small $^{\ast}$ Corresponding author: Kosuke Tatsumura (e-mail: kosuke.tatsumura@toshiba.co.jp)
}
\date{}

\maketitle

\begin{abstract}
Ising machines are specialized computers for finding the lowest energy states of Ising spin models, onto which many practical combinatorial optimization problems can be mapped.
Simulated bifurcation (SB) is a quantum-inspired parallelizable algorithm for Ising problems that enables scalable multi-chip implementations of Ising machines.
However, the computational performance of a previously proposed multi-chip architecture tends to saturate as the number of chips increases for a given problem size because both computation and communication are exclusive in the time domain.
In this paper, we propose a streaming architecture for multi-chip implementations of SB-based Ising machines with full spin-to-spin connectivity.
The data flow in in-chip computation is harmonized with the data flow in inter-chip communication, enabling the computation and communication to overlap and the communication time to be hidden.
Systematic experiments demonstrate linear strong scaling of performance up to the vicinity of the ideal communication limit determined only by the latency of chip-to-chip communication.
Our eight-FPGA (field-programmable gate array) cluster can compute a 32,768-spin problem with a high pipeline efficiency of 97.9\%.
The performance of a 79-FPGA cluster for a 100,000-spin problem, projected using a theoretical performance model validated on smaller experimental clusters, is comparable to that of a state-of-the-art 100,000-spin optical Ising machine.
\end{abstract}

\section{Introduction}
\label{sec:introduction}
Combinatorial optimization problems have many applications in finance \cite{Tatsu_Currency, hidaka2023}, wireless communications \cite{Musen_Kansho, CIM_NOMA}, network infrastructure \cite{Network}, and power distribution \cite{D-wave_PowerGrid}.
Many of them can be mapped onto an energy minimization problem of Ising spin models \cite{Lucas}, which are systems of binary variables having pairwise interactions and individual biases.
The Ising problem is known to be NP-hard (non-deterministic polynomial-time hard) \cite{complexity}.
Many Ising machines, which are special-purpose computers for Ising problems, have been proposed and demonstrated \cite{D-wave,DA,cim10,STATICA,MA,ISCA22,p-bit,SPIM,Eular-SIM}.
Ising machines employ three main approaches: classical annealing, quantum annealing, and dynamical system evolution \cite{ising_compare}.
Simulated annealing (SA) and its derivatives are widely used in classical annealing.

Goto \textit{et~al.} proposed a quantum-inspired simulated bifurcation (SB) algorithm and showed its hardware implementation [hereafter, simulated bifurcation machine (SBM)] \cite{asbm}, which is a type of the Ising machine.
The SBM is derived from the classical counterpart to a quantum bifurcation machine, which is a quantum computer based on the quantum adiabatic theorem \cite{QbM1, QbM2}.
Compared with SA, the SB algorithm has greater parallelism and a higher capability to be accelerated by using parallel computers such as FPGAs (field-programmable gate arrays) and GPUs (graphics processing units).
The FPGA implementations of SB algorithms are remarkable in terms of their quick response times and high energy efficiency \cite{FPL19}.
Tatsumura \textit{et~al.} proposed automated financial trading systems \cite{Tatsu_Currency, tatsumura2023pairstrading, tatsumura2023realtime} that use FPGA-based SB accelerators for detecting trading opportunities and Matsumoto \textit{et~al.} proposed a distance-based clustering method for unevenly distributed data featuring iterative executions of parameter updating and discrete optimization using a general-purpose processor and an FPGA-based SBM (as a low-latency Ising machine) \cite{Clustering_SB}.

Processing of large tasks using multiple chips such as FPGAs or ASICs (application-specific integrated circuits) has been studied for search engines \cite{catapult_v1,catapult_v2}, cryptography \cite{catapult_v2}, N-body simulation \cite{Sano_Nbody,GRAPE-9}, ray tracing \cite{Boku_multi}, artificial intelligence \cite{DFX,AIgean}, and quantum Fourier transform \cite{9798809}.
In addition, an interface for multiple FPGAs similar to the Message Passing Interface has also been proposed \cite{MPIonFPGA}.
The finite die size of a chip determines the number of logic gates per chip, which limits the task size and performance. 
Multi-chip implementations can overcome these limitations.

Several Ising machines using multiple chips have been proposed.
Yamamoto \textit{et~al.} demonstrated a 1.3-Mbit multi-ASIC Ising machine that supported a type of local connectivity called King's graphs \cite{9634769}.
Muthumala and Hariyama reported a multi-FPGA implementation of simulated quantum annealing \cite{8918417}.
Honjo \textit{et~al.} showed a 100,000-spin coherent Ising machine (CIM) using a hybrid system of optics with a 5-km fiber cavity and electronics with multiple FPGAs \cite{cim10}.
Sharma \textit{et~al.} proposed a multi-ASIC Ising machine that combined analog computation and digital inter-chip communication \cite{ISCA22}.
Yamamoto and Kawahara showed a fully digital multi-FPGA Ising machine based on pseudo annealing \cite{YAMAMOTO}.
Kawamura \textit{et~al.} proposed an annealing algorithm called RPA (ratio-controlled parallel annealing) and showed its multi-ASIC implementation \cite{amorphica}.

\begin{figure}[t]
\centering
\includegraphics{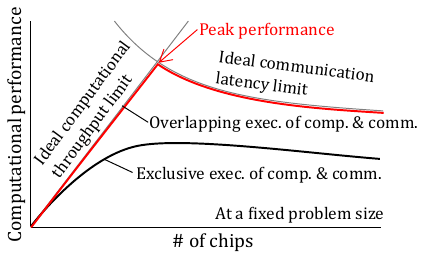}
\caption{
Cluster size (number of chips) dependency of computational performance at a fixed problem size for different cluster architectures featuring the overlapping execution (this work) and exclusive execution (a previous work) of computation and communication. \label{fig:image}
}
\end{figure}

In our previous paper, we reported a multi-chip architecture based on a partitioned version of the SB algorithm to enlarge the number of spins supported and enhance the computational performance \cite{multi_v1}.
In the partitioned SB, all-to-all communication between chips is required for every SB time-evolution step, as in N-body simulations with long-range interaction.
In this architecture, computation and communication are exclusive in the time domain.
Hence, the computational performance starts to saturate when the computation time per SB step decreases to be comparable to the communication time as the number of chips increases for a given problem size (Fig. \ref{fig:image}).
As illustrated in Fig. \ref{fig:image}, if overlapping execution of computation and communication can be realized, the computational performance would then be limited by the communication latency (chip-to-chip latency), not by the communication time (data transfer time).

In this paper, we propose a scalable streaming architecture for SB-based Ising machines with full spin-to-spin connectivity to maximize the peak performance that can be achieved when increasing the cluster size for a given problem size.
In this architecture, the data flow in the in-chip computation is harmonized with the data flow in the inter-chip communication, leading to the overlapping execution of computation and communication.
The performance scales linearly with the number of chips until it is limited only by the communication latency, achieving almost the ideal strong scaling property.
We constructed a theoretical performance model by quantitatively analyzing the internal operation of the cluster architecture and verified the model by systematically comparing the model performance and the measured performance for a series of experimental clusters with various architecture parameters.
We show that the communication time does not affect the computational performance when the computation time per SB step is longer than the communication latency.
We also provide a performance comparison with a state-of-the-art large-scale Ising machine that uses multiple FPGAs.

\section{Cluster Design}
\label{sec:cluster}

\begin{figure*}[t]
\centering
\includegraphics[width=0.80\linewidth]{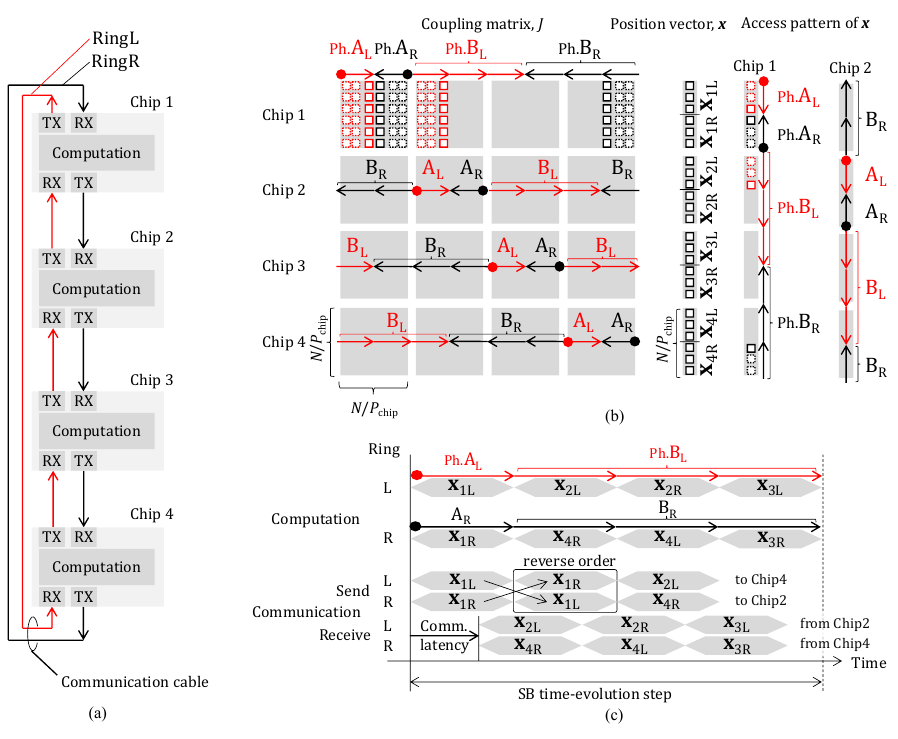}
\caption{
Cluster design (a) Network topology of the cluster. (b) Partition and order of the computation.
(c) Timing chart of each chip (Chip 1 as an example) [the latency of computation and communication logic is ignored for explanatory simplicity].\label{fig:motion}
}
\end{figure*}

We describe the cluster architecture. See the nomenclature in Table \ref{tab:nomenclature} for the abbreviations. 
We designed clusters using a dual-ring communication network shown in Fig. \ref{fig:motion}(a).
Let the two rings be RingR and RingL.
The chips are connected physically in a ring configuration using full-duplex network cables.
The communication between chips that are not directly connected is assisted via transfers by other chips (multi-hop communication).
\begin{table}
\caption{Nomenclature\label{tab:nomenclature}}
\vspace{1em}
\begin{tabularx}{\linewidth}{lX}
$N$ & Number of spin variables of problems or number of SB oscillators in clusters. \\
$P_\mathrm{chip}$ & Cluster size, number of chips.\\
$P_\mathrm{c}$ & Column parallelism, number of processed spin variables per clock cycle per ring at each module.\\
$P_\mathrm{r}$ & Row parallelism, number of spin variables for which each chip is responsible, $=N/P_\mathrm{chip}$.\\
$P_\mathrm{comp}$ & Total parallelism per chip, number of MAC units per chip, $= 2P_\mathrm{r}P_\mathrm{c}$.\\
$N_\mathrm{hop}$ & Number of hops in the all-to-all communication, $= \mathrm{ceil} [(P_\mathrm{chip} - 1)/2]$\\
$N_\mathrm{SBstep}$ & Number of SB steps set by user.\\
$M_\mathrm{step}$ & Number of clock cycles per SB step.\\
$M_\mathrm{compelem}$ & Number of clock cycles to process $N/P_\mathrm{chip}$ elements of positions in each module, characteristic time that determines the operational modes in relation to $\lambda_\mathrm{comm}$, $=N/(2P_\mathrm{chip}P_\mathrm{c})$.\\
$M_\mathrm{commelem}$ & Number of clock cycles to send/receive one subvector (equal to $M_\mathrm{compelem}$).\\
$F_\mathrm{kernel}$ & Clock frequency for computation modules in MHz.\\
$T_\mathrm{step}$ & Processing duration per SB step in $\mu$s.\\
$\lambda_\mathrm{comm}$ & Chip-to-chip communication latency, latency for moving data forward one hop in communication. \\ %, latency from the input to the TX module until the input to the next chip's TX module in cycles.\\
$\lambda_\mathrm{comp}$ & Total latency of operations in TE and MM modules and internal FIFO (first in, first out) in cycles.\\
$\lambda_\mathrm{PHY}$ & Latency from the transmission on TX module to the reception on TE module in the next chip in ns.
\end{tabularx}
\end{table}

The SB algorithm (see Appendix B) introduces interacting oscillators corresponding to the spins in Ising models (see Appendix A), and the entire system corresponds to an oscillator network whose edges represent the couplings between the oscillators. 
The algorithm simulates the time evolution of the oscillator network step-by-step.
In the cluster architecture, the oscillator network is partitioned into multiple subsystems (see Appendix B for the partitioned SB algorithm), each of which corresponds to a chip in the cluster.
Each chip is responsible for updating the positions and momenta of a subset of the oscillators every SB step (the time step for the simulation).
To compute the interaction forces acting on an oscillator, the position data for all the oscillators are needed.

The chips share their position data through the communication performed during the computation of the interaction forces in order to hide the communication time (overlapping execution).
There are two requirements to fully overlap them.
First, the order in which position data is used in the computation should match the order in which they are received in the communication.
If these orders are different, the position data cannot be used in the computation immediately when received, resulting in stalls (temporary halt in computation).
Second, each position datum should be used only once in the computation.
If the data needs to be accessed more than once at different points in time, the second access cannot overlap with the receiving operation.
To satisfy the above conditions, we employed a stream processing, which processes inputs one after another in the same order and at the same throughput in the system.
In addition, the communication throughput should be greater than or equal to the computational throughput to supply inputs for the computation sufficiently in the stream processing.
This overlap can hide the communication time except for initial chip-to-chip communication latency (every SB step).

To hide the initial chip-to-chip communication latency, the computation starts before the arrival of initial data from neighboring chips (every SB step).
At the beginning of each SB step, each chip has only the position data for which the chip is responsible.
Each chip uses these position data before receiving the next ones, preventing computation stalls due to communication latency.

SB algorithm updates the positions and momenta based on the interaction forces every SB step.
The forces are obtained by multiplying a coupling coefficient matrix $J$ with the vector of the positions $\boldsymbol{x}$ [matrix-vector multiplication (MM)].
The matrix $J$ has $N \times N$ elements, which are oscillator-to-oscillator coupling coefficients, where $N$ is the number of oscillators in the cluster. The computation is divided into two parts: MM part, and time-evolution computation part (TE).
TE updates the positions and momenta based on the result of MM.
MM is dominant in terms of the amount of computation; the computational complexity of MM is $\Theta(N^2)$, whereas that of TE is $\Theta(N)$.
Since the computational performance of MM is likely to be limited by memory throughput due to its low arithmetic intensity (one MAC operation per required memory access), our previous work \cite{multi_v1} and this work utilize high memory throughput by using multiple chips and allocating the matrix $J$ to on-chip memories instead of off-chip memory.

Each chip is equally allocated $N/P_\mathrm{chip}$ oscillators. The position data in each chip are further divided into two parts (corresponding to RingR and RingL).
As shown in Fig. \ref{fig:motion}(b), we divide the whole position vector $\boldsymbol{x}$ into $2P_\mathrm{chip}$ subvectors and label them based on the rings and the chips.
For example, Chip 1 updates and stores two subvectors ($\mathrm{\mathbf{x}_{1L}}$ and $\mathrm{\mathbf{x}_{1R}}$).
In accordance with the vector allocation, the coupling coefficient matrix $J$ is also partitioned and allocated equally to the chips.
The size of each submatrix is $(N/P_\mathrm{chip})\times N$.

In every SB step, all the position data must be shared between all the chips (all-to-all communication).
At the beginning of each SB step, each chip has only the two subvectors for which it is responsible.
These subvectors should be sent to all the chips, and the other $2P_\mathrm{chip} - 2$ subvectors should be obtained by the communication.
To enable the multi-hop communication between non-directly connected chips [e.g., Chip 2 and Chip 4 in Fig. \ref{fig:motion}(a)], each chip transfers the received subvectors to its adjacent chips.
To minimize the number of hops in the all-to-all communication, each chip sends the two subvectors for which it is responsible through both rings.
For an example, as shown in Fig. \ref{fig:motion}(c), Chip 1 sends $\mathrm{\mathbf{x}_{1L}}$ and $\mathrm{\mathbf{x}_{1R}}$ to the adjacent chips, Chip 4 and Chip 2, respectively and then sends $\mathrm{\mathbf{x}_{1R}}$ and $\mathrm{\mathbf{x}_{1L}}$ to Chip 4 and Chip 2, respectively [The orders to send the elements in the subvectors, $\mathrm{\mathbf{x}_{1R}}$ and $\mathrm{\mathbf{x}_{1L}}$, are changed (reversed) according to the rings to realize symmetrical operation. Also see the next section for the details.]. Concurrently (after a communication latency, $\lambda_\mathrm{comm}$), Chip 1 starts receiving $\mathrm{\mathbf{x}_{2L}}$ and $\mathrm{\mathbf{x}_{4R}}$ from Chip 2 and Chip 4, respectively. Afterward, Chip 1 sends $\mathrm{\mathbf{x}_{2L}}$ and $\mathrm{\mathbf{x}_{4R}}$ to Chip 4 and Chip 2, respectively.
Eventually, Chip 1 receives six subvectors through communication.

As explained above, to realize overlapping the computation and the communication, their orders should match, and each position datum should be used only once.
We arranged the order of the multiply-accumulate (MAC) operations to fulfill these requirements, as shown in Fig. \ref{fig:motion}(b).
Fig. \ref{fig:motion}(c) shows an example of the timing for the communication and the computation in Chip 1.
In RingL of this example, the subvectors (position data) are used in the order of ``$\mathrm{\mathbf{x}_{1L}}$ $\mathrm{\mathbf{x}_{2L}}$ $\mathrm{\mathbf{x}_{2R}}$ $\mathrm{\mathbf{x}_{3L}}$'', which is the same order as the receive operation except for $\mathrm{\mathbf{x}_{1L}}$ (generated by Chip 1 in the previous SB step).

As already mentioned, to hide the initial chip-to-chip communication latency, the computation should start with the two subvectors for which each chip is responsible, corresponding to phase A (ph.$\mathrm{A_L}$/ph.$\mathrm{A_R}$) in Fig. \ref{fig:motion}(b) and \ref{fig:motion}(c).
These subvectors are available at the beginning of each SB step.
During phase A, each chip can receive and store some inputs for phase B if the communication latency is sufficiently short.
After phase A, the computation proceeds phase B (ph.$\mathrm{B_L}$/ph.$\mathrm{B_R}$), which uses the received position data.
In this way, as shown in the timing chart in Fig. \ref{fig:motion}(c), the computation can be performed without stalling if the communication latency is shorter than the computation time in phase A (corresponding to one hexagon). More details of the timing are provided in Section \ref{sec:scalability}.

Similarly, the stalls while sending caused by the communication latency can be avoided by outputting the position data for which each chip is responsible first.
As shown in Fig. \ref{fig:motion}(c), each chip sends the two subvectors for which it is responsible ($\mathrm{\mathbf{x}_{1L}}$ and $\mathrm{\mathbf{x}_{1R}}$) at the beginning of each SB step.
The communication works without stalling if the communication latency is shorter than the time of sending the two subvectors (corresponding to two hexagons).

\section{Chip Design}
\label{sec:chip}

This section describes the chip-level architecture and discusses its parallelism.

\begin{figure}[t]
\centering
\includegraphics[width=\linewidth-1cm]{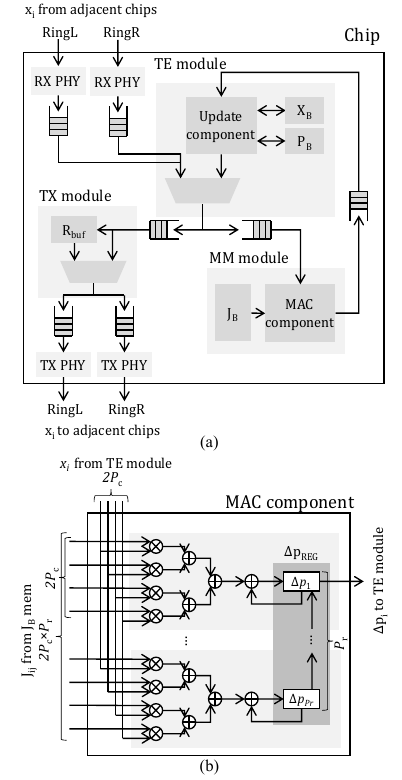}
\caption{
Chip design. (a) Block diagram for each chip.
(b) Implementation details of MAC component.
$\mathrm{\Delta p_\mathrm{REG}}$ serves as both the accumulators of the MAC units and the shift register for output.
\label{fig:arch}
}
\end{figure}

Each chip has four functions: MM, TE, receive operations, and send operations.
These functions are assigned to three modules as shown in Fig. \ref{fig:arch}(a).
The MM computation is assigned to the MM module.
The receive operations and TE computation are assigned to the TE module.
The send operations are assigned to the TX module.
There are three on-chip memories: $\mathrm{J_B}$, $\mathrm{X_B}$, and $\mathrm{P_B}$.
They store the coupling coefficient matrices [$N \times (N/P_\mathrm{chip})$ elements], the position data ($N/P_\mathrm{chip}$ elements), and the momentum data ($N/P_\mathrm{chip}$ elements), respectively.
There are also queues for the RX PHYs, the TX PHYs, and between the modules.

The computation and the communication operate at the same processing throughput (stream processing).
The processing throughput is determined by the parallelism of the MM module.
The MM module calculates and accumulates the interaction forces acting on the $N/P_\mathrm{chip}$ oscillators for which each chip is responsible.
The MM module consists of the $\mathrm{J_B}$ memory and a MAC component as shown in Fig. \ref{fig:arch}(a).
Fig. \ref{fig:arch}(b) shows the internal architecture of the MAC component.
The total parallelism (the number of MAC units, $P_\mathrm{comp}$) in the MM module is determined to be maximized within the limits of the hardware logic resources on a chip.
As described in the previous section, the computation should use the received position data only once; therefore, all $N/P_\mathrm{chip}$ MAC operations using a received position datum are performed together in parallel.
The row parallelism ($P_\mathrm{r}$) is determined to be $N/P_\mathrm{chip}$.
Since the total parallelism ($P_\mathrm{comp}$) and the row parallelism ($P_\mathrm{r}$) are already determined, the column parallelism is determined to be $P_\mathrm{comp} / P_\mathrm{r}$ (let $2P_\mathrm{c}$ be it, $P_\mathrm{c}$ for each ring), which is equal to the input throughput of the MM module.
All the processing throughputs of the position data for the computation and the communication are adjusted with $P_\mathrm{c}$ for each ring.
The resource usage has been optimized because there is no wasted throughput.
For the details of the resource usage, see Section S2 in Supplementary Material.

To realize the overlap, the size of $\Delta p_\mathrm{REG}$ to store the intermediate values of the MAC component [Fig. \ref{fig:arch}(b)] should be $P_\mathrm{r}$, since the $P_\mathrm{r}$ interaction forces of the oscillators for which each chip is responsible are computed in parallel.
When the number $P_\mathrm{r}$ becomes large (this is the case for this work, see Table \ref{tab:main_table}), aggregating the register outputs is difficult to implement with a selector.
Therefore, we use a shift-register structure for $\Delta p_\mathrm{REG}$, which enables choosing a large architectural parameter of $P_\mathrm{r}$ like $P_\mathrm{r}=8,192$.

TX module is designed to prevent propagation of stalls to adjacent chips.
As noted above, the order in which the subvector elements (position data) are sent depends on the rings.
At the beginning of the SB step, each chip sends the two subvectors for which it is responsible through both rings.
Each subvector is sent earlier through one ring and later through the other ring.
For example, as shown in Fig. \ref{fig:motion}(c), $\mathrm{\mathbf{x}_{1R}}$ is sent earlier through RingR and later through RingL.
The element order of the subvector sent later is the reverse order of that sent earlier through the opposite ring.
Therefore, $\mathrm{R_{buf}}$ in Fig. \ref{fig:arch}(a) stores the subvector ($N/P_\mathrm{chip}$ elements of position data per ring) during the earlier sending and later autonomously sends them in the reverse order through the opposite ring.
By separating the TE module (receive operation) and the TX module (send operation), the control flows of receiving and sending are separated.
Thus, the chain propagation of stalls in the communication is prevented as much as possible.

\section{Scalability}
\label{sec:scalability}

In this section, we introduce a performance model based on our quantitative analysis.
This model clarifies the factors that limit the computational performance.
Using this model, we discuss the scalability of the proposed architecture.
In the next section, we experimentally confirm that the model gives a precise estimation of the performance. 

The scalability of cluster computing is often discussed in terms of weak scaling and strong scaling.
Weak scaling is the dependence of performance on cluster size with a fixed ratio of the amount of computation to cluster size (with a fixed amount of computation per chip regardless of cluster size).
Strong scaling is the dependence of performance on cluster size with a fixed amount of computation (the amount of computation per chip decreases with cluster size).
The performance of the proposed architecture is limited by its computational throughput (determined by the parallelism), the communication latency, or both.
We distinguish these three cases as operational modes and discuss their performance (the throughput of the processed MAC operations including overheads).

\subsection{Cycle-level performance model}

Before discussing the performance directly, the number of hardware clock cycles should be discussed.
The total number of clock cycles per execution is determined by the number of SB steps ($N_\mathrm{SBstep}$) and the number of clock cycles per SB step ($M_\mathrm{step}$).
We here discuss $M_\mathrm{step}$, while $N_\mathrm{SBstep}$ is determined by users from the viewpoint of solution quality (the quality of solution is improved as $N_\mathrm{SBstep}$ is increased in general \cite{asbm}).

\begin{figure}[t]
\centering
\includegraphics[width=\linewidth-5mm]{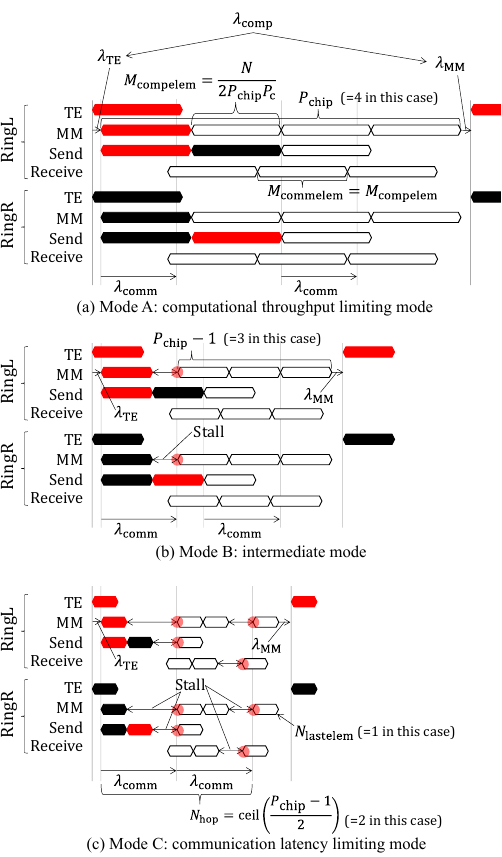}
\caption{
Timing chart of a chip for each mode.
Each of the hexagons corresponds to the duration of inputting a subvector.
(a) Mode A: computational throughput limiting mode. (b) Mode B: intermediate mode. (c) Mode C: communication latency limiting mode.
The red and black filled hexagons represent the duration of inputting a subvector computed by the chip itself.
The non-filled hexagons represent the duration of inputting a subvector received from the other chips. 
\label{fig:timing}
}
\end{figure}

Fig. \ref{fig:timing} shows the timing chart of an SB step for each mode.
$\lambda_\mathrm{comp}$ is the latency of computational modules per SB step (the sum of $\lambda_\mathrm{TE}$ and $\lambda_\mathrm{MM}$).
$\lambda_\mathrm{comm}$ is the latency for moving data forward one hop in communication (the sum of $\lambda_\mathrm{TE}$, $\lambda_\mathrm{TX}$, and $\lambda_\mathrm{PHY}$).
$\lambda_\mathrm{TE}$, $\lambda_\mathrm{MM}$, and $\lambda_\mathrm{TX}$ are the latencies for passing through TE, MM, and TX modules respectively.
$\lambda_\mathrm{PHY}$ is the latency from the transmission on the TX module to the reception on the TE module in the next chip.
$N_\mathrm{hop}$ is the number of hops required in the all-to-all communication in each SB step: $N_\mathrm{hop} = \mathrm{ceil} [(P_\mathrm{chip} - 1)/2]$.

Each hexagon in Fig. \ref{fig:timing} represents the number of cycles to input a subvector into each module.
All modules are fully pipelined so that no stall occurs while processing subvectors.
Therefore, the lengths of the hexagons are always constant.
The number of cycles for inputting a subvector to the MM module is given by
\begin{align}
M_\mathrm{compelem} = \cfrac{N}{2P_\mathrm{chip} P_\mathrm{c}}.
\end{align}
$M_\mathrm{compelem}$ is a characteristic time for this architecture because it determines the operational modes.
Not only the MM module but also the TE module and the TX module have the same processing throughput.
In addition, the number of cycles for sending or receiving a subvector ($M_\mathrm{commelem}$) is equal to $M_\mathrm{compelem}$.

There are two ideal numbers of cycles per SB step.
The ideal number of cycles for the computation ($M^\mathrm{ideal}_\mathrm{step}$) is $N^2/(P_\mathrm{chip}P_\mathrm{comp})$, corresponding to the number of requested MAC operations ($N^2$) divided by the total number of MAC units in a cluster ($P_\mathrm{chip}P_\mathrm{comp}$).
It is related to the $M_\mathrm{compelem}$ with $N^2/(P_\mathrm{chip}P_\mathrm{comp}) = P_\mathrm{chip} M_\mathrm{compelem}$.
On the other hand, the ideal number of cycles for the all-to-all communication per SB step is $N_\mathrm{hop} \lambda_\mathrm{comm}$, which is the product of the number of hops ($N_\mathrm{hop}$) and the chip-to-chip latency ($\lambda_\mathrm{comm}$).
The actual number of execution cycles per SB step (corresponding to $M_\mathrm{step}$ in the following) is larger than both of these two ideal numbers of cycles.

$M_\mathrm{step}$ (the number of cycles per SB step) is expressed for three operational modes: computational throughput limiting mode (Mode A), intermediate mode (Mode B), and communication latency limiting mode (Mode C).
The operational mode of an cluster is determined by the relationship between $\lambda_\mathrm{comm}$ (communication latency) and $M_\mathrm{compelem}$ (characteristic time).
The theoretical performance model of the proposed cluster architecture is defined by Eqs. (\ref{eq:comp}), (\ref{eq:multi}), and (\ref{eq:comm}) as following.
\begin{flalign}
& M_\mathrm{step}= & \notag
\end{flalign}
\begin{subnumcases}{}
P_\mathrm{chip} M_\mathrm{compelem} + \lambda_\mathrm{comp} \label{eq:comp} \\
\hspace{0.3cm} \text{for\ Mode A:\ } \lambda_\mathrm{comm} \leq M_\mathrm{compelem}, \notag \\
(P_\mathrm{chip}-1) M_\mathrm{compelem} + \lambda_\mathrm{comm} + \lambda_\mathrm{comp} \label{eq:multi} \\
\hspace{0.3cm} \text{for\ Mode B:\ } M_\mathrm{compelem} < \lambda_\mathrm{comm} \leq 2M_\mathrm{compelem}, \notag \\
N_\mathrm{hop} \lambda_\mathrm{comm} + N_\mathrm{lastelem} M_\mathrm{compelem} + \lambda_\mathrm{comp} \label{eq:comm} \\
\hspace{0.3cm} \text{for\ Mode C:\ } 2M_\mathrm{compelem} < \lambda_\mathrm{comm}, \notag
\end{subnumcases}
\begin{align}
\intertext{where}
N_\mathrm{lastelem} =
\begin{cases*}
1 & if $P_\mathrm{chip}$ is even, \\
2 & if $P_\mathrm{chip}$ is odd.
\end{cases*}
\end{align}
Note that $N_\mathrm{hop}$ is $P_\mathrm{chip}/2$ if $P_\mathrm{chip}$ is even.

The factor that determines $M_\mathrm{step}$ changes depending on the mode.
In Mode A (computational throughput limiting mode), $M_\mathrm{step}$ is expressed by Eq. (\ref{eq:comp}); it is the sum of the number of cycles for a computation ($P_\mathrm{chip} M_\mathrm{compelem}$) and the overhead of $\lambda_\mathrm{comp}$.
Note that $M_\mathrm{step}$ is not affected by the communication because the communication overlaps completely with the computation. 
In Mode B (intermediate mode), a chunk of stalls occurs in MM module due to stalls in the receive operations [see Fig. \ref{fig:timing}(b)].
Equation (\ref{eq:multi}) can be regarded as replacing some of cycles in Eq. (\ref{eq:comp}) with the communication latency $\lambda_\mathrm{comm}$.
$M_\mathrm{step}$ is affected by both the computational throughput and the communication latency.
In Mode C (communication latency limiting mode), the chip-to-chip communication latency $\lambda_\mathrm{comm}$ is larger than $2M_\mathrm{compelem}$, and stalls in sending propagates to all chips in a cluster.
Equation (\ref{eq:comm}) is the sum of the ideal number of cycles for all-to-all communication ($N_\mathrm{hop} \lambda_\mathrm{comm}$) and overhead ($N_\mathrm{lastelem} M_\mathrm{compelem} + \lambda_\mathrm{comp}$).
$M_\mathrm{step}$ is determined mainly by the communication latency.

Memory throughput and communication throughput do not affect $M_\mathrm{step}$ in any modes, because the implemented hardware has sufficient throughput for both (for the discussion of these throughputs, see Sections S4 and S5 in Supplementary Material).

\subsection{Weak scaling and strong scaling}
We define the performance for discussing scalability.
In this paper, the performance is defined by the number of MAC operations per second, since the MAC operations in MM are the dominant computations.
The time for an SB step is expressed as $T_\mathrm{step} = M_\mathrm{step}/F_\mathrm{kernel}$, where $F_\mathrm{kernel}$ is the clock frequency.
Then, the performance (throughput of MAC operations) can be written as 
\begin{align}
\mathrm{Performance} = \cfrac{N(N-1)}{T_\mathrm{step}} = \cfrac{N(N-1)F_\mathrm{kernel}}{M_\mathrm{step}}. \label{eq:perf}
\end{align}
Here, $N(N-1)$ is the number of MAC operations performed per SB step.

We next discuss the scalability from the points of weak scaling ($N$ and $P_\mathrm{chip}$ increase proportionally) and strong scaling ($P_\mathrm{chip}$ increases with a fixed $N$).
Here, we assume that $P_\mathrm{comp}$ ($= 2P_\mathrm{c}P_\mathrm{r}$, the number of MAC units per chip) is constant because it is approximately determined by the resource on a chip.
Note that, in our definition of weak scaling, the number of MAC operations per chip increases in proportion to $P_\mathrm{chip}$, because the computational complexity of MM is $\Theta(N^2)$.

In the case of weak scaling, the number of oscillators per chip $N/P_\mathrm{chip}$ ($=P_\mathrm{r}$) and $P_\mathrm{c}$ ($=P_\mathrm{comp}/2P_\mathrm{r}$) are constant.
Therefore, the operational mode is not changed, because both $M_\mathrm{compelem}$ and $\lambda_\mathrm{comm}$ are constants.
The $M_\mathrm{step}$ in Eqs. (\ref{eq:comp}) and (\ref{eq:multi}) are linear with respect to $P_\mathrm{chip}$.
The $M_\mathrm{step}$ in Eq. (\ref{eq:comm}) is also linear with respect to $P_\mathrm{chip}$ under assuming $P_\mathrm{chip}$ is even [First term of Eq. (\ref{eq:comm}) is ($P_\mathrm{chip} \lambda_\mathrm{comm})/2$, and second term and third term of the equation are constants in this assumption].
Hence, $M_\mathrm{step}$ increases linearly as $P_\mathrm{chip}$ increases.
According to Eq. (\ref{eq:perf}), the performance is a linear function of $P_\mathrm{chip}$, disregarding the constant terms in $M_\mathrm{step}$.
This scaling characteristic is ideal.

In the case of strong scaling, since $N$ is fixed, $N/P_\mathrm{chip}$ ($=P_\mathrm{r}$) decreases and $P_\mathrm{c}$ ($=P_\mathrm{comp}/2P_\mathrm{r}$) increases as $P_\mathrm{chip}$ increases.
Hence, $M_\mathrm{compelem}$ decreases as $P_\mathrm{chip}$ increases, and the mode transitions in the order of Mode A, Mode B, and Mode C.
In Mode A and Mode B, the performance increases with increasing $P_\mathrm{chip}$.
In Mode C, the performance is limited by the communication latency and degrades as $P_\mathrm{chip}$ increases.
Therefore, the peak performance is achieved at the mode transition point from Mode B to Mode C.
For the details of performance trends in Mode B and Mode C, see Section S6 in Supplementary Material.

Figure \ref{fig:image} illustrates the strong scaling property of the proposed architecture when $P_\mathrm{chip}$ varies.
The gray lines represent the ideal computational throughput limit calculated from the ideal number of cycles for the computation [$N^2/(P_\mathrm{chip}P_\mathrm{comp})$] and the ideal communication latency limit calculated from the ideal number of cycles for the all-to-all communication ($N_\mathrm{hop} \lambda_\mathrm{comm}$). See Section S7 in Supplementary Material for detailed information of the ideal performance.
The red line corresponds to the theoretical model performance calculated from $M_\mathrm{step}$ [Eqs. (\ref{eq:comp}), (\ref{eq:multi}), and (\ref{eq:comm})].
Thus, Fig. \ref{fig:image} illustrates how close the model performance is to the ideal limit.
In the next section, we quantitatively show how close the measured performance in experiments is to the model performance and to the ideal limits.
A similar discussion for general purpose processors has been given by the Roofline model \cite{Roofline}, which varies operational intensity (operations per byte of DRAM traffic) instead of $P_\mathrm{chip}$.

\section{Performance}
\label{sec:performance}

\begin{table*}[!tb]
\caption{
Measured and theoretical model performance of the proposed clusters for SBMs.
Design IDs are assigned to each of the synthesized circuits.
Cluster IDs are assigned to each combination of design and number of chips in a cluster.
``Theory (Syn.)'' is calculated based on the architecture parameters of the synthesized hardware.
``Theory (Typ.)'' is calculated based on the typical parameters shown in the column on the far right.
See the nomenclature in Table \ref{tab:nomenclature} for the definitions of other symbols.
\label{tab:main_table} }

\resizebox{\linewidth}{!}{
\begin{tabular}{l|l|ccc|ccc|ccc|c|c}
\hline 
\multicolumn{2}{l|}{Design ID} & \multicolumn{3}{c|}{S1K} & \multicolumn{3}{c|}{S2K} & \multicolumn{3}{c|}{S4K} & S8K & Typical \\ \hline
\multicolumn{2}{l|}{$N/P_\mathrm{chip} (=P_\mathrm{r})$}
& \multicolumn{3}{c|}{1,024} & \multicolumn{3}{c|}{2,048} & \multicolumn{3}{c|}{4,096} & 8,192 & -- \\ \hline
\multicolumn{2}{l|}{Operational mode of the model}
& \multicolumn{3}{c|}{Mode C: communication } & \multicolumn{3}{c|}{Mode B:} & \multicolumn{4}{c|}{Mode A: computational} & \\
\multicolumn{2}{l|}{} & \multicolumn{3}{c|}{latency limiting mode } & \multicolumn{3}{c|}{intermediate mode} & \multicolumn{4}{c|}{throughput limiting mode} & -- \\ \hline
\multicolumn{2}{l|}{$M_\mathrm{compelem}$ ($=N/(2P_\mathrm{chip}P_\mathrm{c})$)} & \multicolumn{3}{c|}{} & \multicolumn{3}{c|}{} & \multicolumn{3}{c|}{} & & \\
\multicolumn{2}{l|}{(cycles) } & \multicolumn{3}{c|}{32} & \multicolumn{3}{c|}{128} & \multicolumn{3}{c|}{512} & 2,048 & -- \\ \hline
\multicolumn{2}{l|}{$\lambda_\mathrm{comm}$ (cycles)} & \multicolumn{3}{c|}{177} & \multicolumn{3}{c|}{181} & \multicolumn{3}{c|}{177} & 167 & 177 \\ \hline \hline
\multicolumn{2}{l|}{$P_\mathrm{c}$} & \multicolumn{3}{c|}{16} & \multicolumn{3}{c|}{8} & \multicolumn{3}{c|}{4} & 2 & -- \\ \hline
\multicolumn{2}{l|}{$P_\mathrm{comp} (=2P_\mathrm{r}P_\mathrm{c})$}
& \multicolumn{10}{c|}{32,768} & 32,768 \\ \hline
\multicolumn{2}{l|}{$\lambda_\mathrm{comp}$ (cycles)} & \multicolumn{3}{c|}{81} & \multicolumn{3}{c|}{80} & \multicolumn{3}{c|}{87} & 101 & 85 \\ \hline
\multicolumn{2}{l|}{$F_\mathrm{kernel}$(MHz)}& \multicolumn{3}{c|}{281} & \multicolumn{3}{c|}{301} & \multicolumn{3}{c|}{303} & 275 & 293 \\ \hline
\multicolumn{2}{l|}{Comm. throughput (Gbps)}
& \multicolumn{3}{c|}{72} & \multicolumn{3}{c|}{39} & \multicolumn{3}{c|}{19} & 9 & -- \\ \hline \hline
\multicolumn{2}{l|}{Cluster ID} & \#1 & \#2 & \#3 & \#4 & \#5 & \#6 & \#7 & \#8 & \#9 & \#10 & -- \\ \hline
\multicolumn{2}{l|}{$N$} & 2,048 & 4,096 & 8,192 & 4,096 & 8,192 & 16,384 & 8,192 & 16,384 & 32,768 & 16,384 & -- \\ \hline
\multicolumn{2}{l|}{$P_\mathrm{chip}$} & 2 & 4 & 8 & 2 & 4 & 8 & 2 & 4 & 8 & 2 & -- \\ \hline
\multicolumn{2}{l|}{$N_\mathrm{hop} (=\lceil (P_\mathrm{chip}-1)/2 \rceil)$}
& 1 & 2 & 4 & 1 & 2 & 4 & 1 & 2 & 4 & 1 & -- \\ \hline
& $M_\mathrm{step}$ (cycles) & 290 & 467 & 820 & 391 & 647 & 1,160 & 1,111 & 2,135 & 4,183 & 4,197 & -- \\ \cline{2-13}
Measured & Performance (GMAC/s) & 4,062 & 10,093 & 22,994 & 12,912 & 31,217 & 69,650 & 18,300 & 38,094 & 77,775 & 17,588 & -- \\ \cline{2-13}
& $T_\mathrm{step}$ ($\mathrm{\mu}$s) & 1.0 & 1.7 & 2.9 & 1.3 & 2.1 & 3.9 & 3.7 & 7.0 & 13.8 & 15.3 & -- \\ \cline{2-13}
& Efficiency (\%) & 22.1 & 27.4 & 31.2 & 65.5 & 79.1 & 88.3 & 92.2 & 95.9 & 97.9 & 97.6 & -- \\ \hline
Theory & $M_\mathrm{step}$ (cycles) & 290 & 467 & 821 & 389 & 645 & 1,157 & 1,111 & 2,135 & 4,183 & 4,197 & -- \\ \cline{2-13}
(Syn.) & Performance (GMAC/s) & 4,062 & 10,093 & 22,966 & 12,979 & 31,314 & 69,831 & 18,300 & 38,094 & 77,775 & 17,588 & -- \\ \hline
Theory & $M_\mathrm{step}$ (cycles) & 294 & 471 & 825 & 390 & 646 & 1,158 & 1,109 & 2,133 & 4,181 & 4,181 & -- \\ \cline{2-13}
(Typ.) & Performance (GMAC/s) & 4,178 & 10,434 & 23,831 & 12,601 & 30,434 & 67,916 & 17,728 & 36,871 & 75,244 & 18,811 & -- \\ \hline
\end{tabular}
}
\end{table*}

\begin{figure*}[!tb]
\centering
\includegraphics[width=\linewidth]{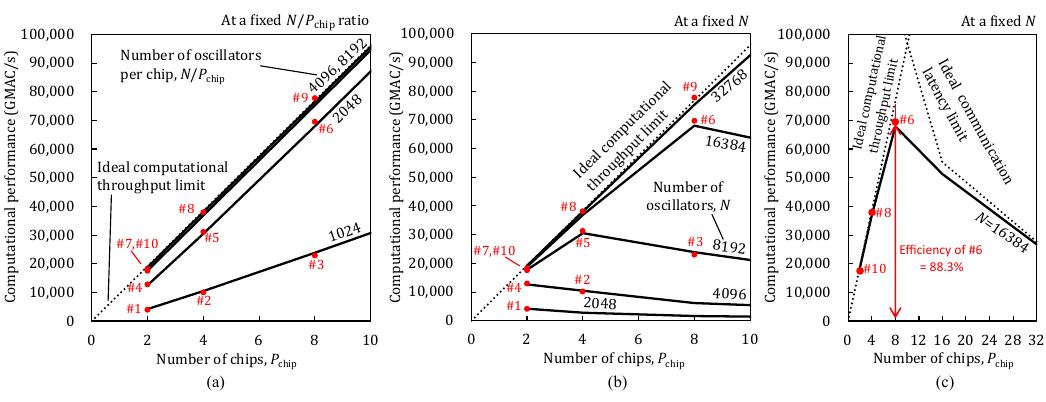}
\caption{
Measured and theoretical-model performance of the proposed clusters for SBMs.
The red plots represent the measured performance.
The solid lines illustrate the model performance.
The dotted lines correspond to the ideal limits determined by computational throughput and communication latency.
(a) Weak scaling characteristics. Increase the number of chips ($P_\mathrm{chip}$) and the problem size ($N$) in the same proportion [the numbers of oscillators per chip $N/P_\mathrm{chip}$ are fixed].
(b) Strong scaling characteristics. Increase the number of chips ($P_\mathrm{chip}$) at a fixed problem size ($N$) [the numbers of oscillators ($N$) are fixed].
(c) Comparison of measured performance and ideal limits at $N = 16,384$.
\label{fig:result}
}
\end{figure*}

We demonstrate the high scalability through experiment results.
We used two types of FPGAs: Stratix10 FPGAs (14-nm process) and Arria10 FPGAs (20-nm process).
For further details of the FPGAs, see Appendix C and Sections S1 and S2 in Supplementary Material.
In this section, we mainly discuss the measurement results for the Stratix 10 FPGA implementation, and then compare the proposed architecture with the results of a previous work \cite{multi_v1} using the same Arria10 FPGA boards.
We used MAX-CUT problems with node numbers equal to the numbers of oscillators in clusters ($N$).

Table \ref{tab:main_table} shows the evaluation results for the clusters with Stratix10 FPGAs.
We evaluated 10 clusters (distinguished by Cluster IDs: \#1-\#10) with four circuit designs (identified by Design IDs: S1K, S2K, S4K, S8K).
Different designs have different numbers of oscillators for which each chip is responsible ($N/P_\mathrm{chip}$).
The number of oscillators in the cluster ($N$) increases as the number of chips ($P_\mathrm{chip}$) increases.
For example, the clusters of 8,192 (\#7), 16,384 (\#8), and 32,768 (\#9) oscillators can consist of 2, 4, and 8 chips with S4K design respectively.
There are some clusters that have the same number of oscillators ($N$); for example, both \#8 (4 chips with S4K design) and \#10 (2 chips with S8K design) have 16,384 oscillators.
The efficiency in Table \ref{tab:main_table} is defined as 
\begin{align}
\mathrm{Efficiency} = \cfrac{M_\mathrm{step}^\mathrm{ideal}}{M_\mathrm{step}} = \cfrac{N^2}{P_\mathrm{comp}P_\mathrm{chip}}\cdot\cfrac{1}{M_\mathrm{step}}.
\end{align}
Note that $M^\mathrm{ideal}_\mathrm{step}=P_\mathrm{chip}M_\mathrm{compelem}=N^2/(P_\mathrm{comp}P_\mathrm{chip})$ is the ideal number of clock cycles for the computation.

The measured results are also shown in Fig. \ref{fig:result} (See Section S3 in Supplementary Material for the measurement details).
Fig. \ref{fig:result}(a) and \ref{fig:result}(b) shows the weak scaling property and the strong scaling property of the architecture, respectively.
The red plots are the measured performance shown in Table \ref{tab:main_table}.
The black solid lines are given by the model with the typical parameters shown in the rightmost column in Table \ref{tab:main_table} (introduced to simplify the discussion).
The dotted lines show the ideal computational throughput limit and the ideal communication latency limit with the typical parameters.

Fig. \ref{fig:result}(a) shows the high efficiency of the proposed architecture at Mode A.
The figure shows the measured performance and the theoretical model performance (GMAC/s) as functions of the number of chips ($P_\mathrm{chip}$) with fixed numbers of oscillators per chip ($N/P_\mathrm{chip}$).
The clusters with $N/P_\mathrm{chip} = 8,192$ (S8K design, \#10), $4,096$ (S4K design, \#7--\#9) are at Mode A.
Their performance is on the ideal limit because the overhead $\lambda_\mathrm{comp}$ is sufficiently small compared with $M_\mathrm{step}$.
For an example, the cluster \#9 (S4K design) achieved the highest pipeline efficiency of 97.9\% and the highest performance of 77,775 GMAC/s in our experiments.
The clusters \#4--\#6 are Mode B, and the clusters \#1--\#3 are Mode C.

Fig. \ref{fig:result}(a) also shows the linearity of the weak scaling property of the proposed architecture.
The model performance (the black lines) in Fig. \ref{fig:result}(a) is linear with respect to $P_\mathrm{chip}$ at any modes (as discussed in the previous section).
We confirmed that the model has provided precise estimations of performance, i.e., the measured performance (the red plots) aligns with the model performance (the black lines).
The slopes of the black lines correspond to the coefficient of $P_\mathrm{chip}$ in Eqs. (\ref{eq:comp}), (\ref{eq:multi}), and (\ref{eq:comm}).
The slope of $N/P_\mathrm{chip}=1024$ (Mode C) is flatter (smaller) than those of Mode A and Mode B, because the coefficient of $P_\mathrm{chip}$ in Eq. (\ref{eq:comm}) is larger than those in Eqs. (\ref{eq:comp}) and (\ref{eq:multi}) due to the communication latency.
The y-intercepts of the black lines represent the constant terms in the equations.
The y-intercept of $N/P\mathrm{chip}=2048$ (Mode B) is different from those of Mode A, because the overhead (a constant term) is larger due to the stalls caused by the initial chip-to-chip communication latency.

Fig. \ref{fig:result}(b) shows the performance (GMAC/s) as a function of the numbers of chips ($P_\mathrm{chip}$) for a fixed number of oscillators ($N$).
The plots are the same as those in Fig. \ref{fig:result}(a), whereas the black lines are different. Fig. \ref{fig:result}(b) shows the strong scaling property of the proposed architecture.
Here, we consider the case of $N = 16,384$ as a typical example. As Table \ref{tab:main_table} shows, when using two chips (\#10) or four chips (\#8), the performance is limited by computational throughput, and the pipeline efficiency is almost ideal.
The performance improvement continues up to eight chips (\#6).

We discuss the limit of strong scaling.
Fig. \ref{fig:result}(c) shows the measured and model performance at $N = 16,384$ and the ideal limits.
The experiment results show the performance improvement along with the ideal computational throughput limit from 2 chips (\#10) to 8 chips (\#6).
At \#6 (eight chips), the experimental performance reaches 69,650 GMAC/s, which is the peak of strong scaling at $N = 16,384$.
This performance is 6.41 times higher than that in our previous work \cite{multi_v1}.
Between 8 chips and 16 chips, the operational mode changes from Mode B to Mode C (communication latency limiting mode), and the performance starts to degrade.
As shown in Fig. \ref{fig:result}(c), the estimated performance with 16 chips is located at the vicinity of the ideal communication latency limit ($N_\mathrm{hop}\lambda_\mathrm{comm}$).
As we can see here, the transition from the ideal computational throughput limit to the ideal communication latency limit is abrupt.
For further performance improvement, technologies that reduce the communication latency or reduce the number of hops (by changing the network topology) are needed.

In the experiments, S2K design has always given the best performance for a given number of oscillators in the clusters $N$.
The performance peaks at the mode transition point from Mode B to Mode C.
The condition for the transition is $\lambda_\mathrm{comm} = 2M_\mathrm{compelem}$.
As shown in Table \ref{tab:main_table}, S2K design is close to this condition: $\lambda_\mathrm{comm} = 181$ (cycles) and $M_\mathrm{compelem} = 128$ (cycles). In addition, the ideal $N/P_\mathrm{chip}$ that maximize the performance can be calculated.
Under the assumption of $\lambda_\mathrm{comm} = 177$ (cycles) and $P_\mathrm{comp} = 32,768$ (typical parameters), the optimal $N/P_\mathrm{chip}$ is given by $\sqrt{P_\mathrm{comp}\lambda_\mathrm{comm}/2}= 1,703$ (close to S2K design).

The proposed architecture provides flexible options for performance and cost.
We have shown the optimal design for performance, however, such design requires many chips relatively.
This architecture also can provide options that use smaller clusters to meet required performance at lower cost.
For example, S4K design can realize clusters with the same number of oscillators but half the number of chips compared with S2K design.

\begin{figure}[t]
\centering
\includegraphics[width=\linewidth]{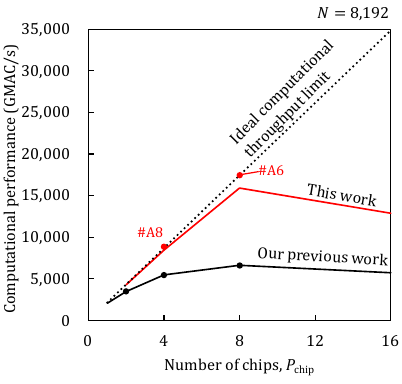}
\caption{
Performance comparison between our previous work \cite{multi_v1} and this work implemented on the same Arria10 FPGA boards for a MAX-CUT problem with 8,192 nodes (oscillators). Red and black plots show the measured performance.
Solid red and black lines show, respectively, the theoretical performance of this architecture with typical parameters and the simulated performance of previous architecture.
For both the ideal computational throughput limit and the theoretical performance, we assume that FPGAs are operating at 266 MHz.
The measured performance on an eight-chip cluster in this work is slightly higher than the theoretical performance because of the difference in clock frequency. \label{fig:arch_diff}
}
\end{figure}

The difference between our previous work \cite{multi_v1} and the present work exists in the strong scaling property.
Fig. \ref{fig:arch_diff} shows a performance comparison between these two works using the same Arria10 FPGA boards.
For the details of the implementation and the performance, see Section S1 in Supplementary Material.
In the previous architecture, as $P_\mathrm{chip}$ has increased, the gap between the performance and the ideal computational throughput limit has become larger, and the clusters have gradually transitioned to the performance degradation phase.
In the proposed architecture, the performance has scaled linearly up to eight chips ($P_\mathrm{chip} = 8$), achieving a performance improvement of 2.63-fold at $P_\mathrm{chip} = 8$.
A better performance has been observed even in the performance degradation phase ($P_\mathrm{chip} > 8$), mainly because the number of hops in the all-to-all communication ($N_\mathrm{hop}$) has been reduced.

\section{Comparison with a State-of-the-Art Ising Machine}
\label{sec:comp_cim}

\begin{figure*}[t]
\centering
\includegraphics[width=\linewidth]{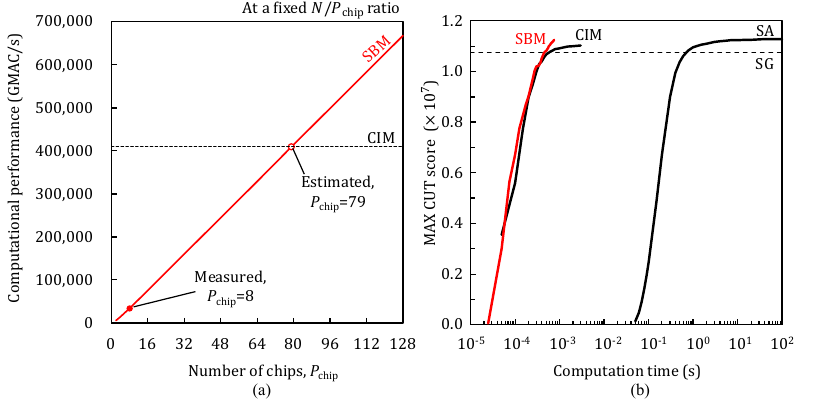}
\caption{
Comparison of the proposed SB-based cluster with the CIM and the SA on a 100,000-node problem. (a) Comparison of computational throughput. (b) Comparison of time-to-target (computation time to reach the target MAX-CUT score) \label{fig:cim}
}
\end{figure*}

We compare the proposed SB-based architecture with a state-of-the-art Ising machine \cite{cim10}, i.e., the 100,000-spin coherent Ising machine (CIM) for 100,000-node MAX-CUT problems.

We designed and synthesized another circuit named S1280 design ($N/P_\mathrm{chip}=1,280$).
It has a submatrix $J$ with the size of $1,280 \times 100,000$ on the single FPGA (Stratix10), enabling a 79-chip cluster to support 100,000-node problems.
The functionality of S1280 design has been validated with an eight-chip experimental cluster (\#11) for 10,240-node MAX-CUT problems, as shown in Table \ref{tab:cim}.
The measured performance is 33,787 GMAC/s, which matches to the model performance (34,418 GMAC/s).
Note that the architectural parameters for this design was adjusted in order to accommodate the large matrix within the single FPGA.

The performance of a 79-chip cluster (\#12) has been projected by the theoretical model with the architectural parameters on Table \ref{tab:cim}.
As shown in Table 3 and Fig. \ref{fig:cim}(a), the projected performance is 408,948 GMAC/s (25.0 $\mu$s per SB step), which is almost the same as the performance of the CIM (408,873 GMAC/s) calculated from 593 $\mu$s for 24 circulations (a circulation of the CIM corresponds to an SB step).

\begin{table}[t]
\caption{Measured and projected performance on the proposed cluster for 100,000-node MAX-CUT problems.
The efficiency is calculated based on the reduced number of MAC units.\label{tab:cim}}
\resizebox{\linewidth}{!}{
\begin{tabular}{l|l|p{0.15\columnwidth}<{\centering}p{0.15\columnwidth}<{\centering}}
\hline
\multicolumn{2}{l|}{Design ID} & \multicolumn{2}{c}{S1280} \\ \hline
\multicolumn{2}{l|}{$N/P_\mathrm{chip} (=P_\mathrm{r})$} & \multicolumn{2}{c}{1,280} \\ \hline
\multicolumn{2}{l|}{Operational mode of the model} & \multicolumn{2}{c}{Mode C: communication} \\
\multicolumn{2}{l|}{} & \multicolumn{2}{c}{latency limiting mode} \\ \hline
\multicolumn{2}{l|}{$M_\mathrm{compelem}$ ($=N/(2P_\mathrm{chip}P_\mathrm{c})$) (cycles)} & \multicolumn{2}{c}{80} \\ \hline
\multicolumn{2}{l|}{$\lambda_\mathrm{comm}$ (cycles)} & \multicolumn{2}{c}{174} \\ \hline \hline
\multicolumn{2}{l|}{$P_\mathrm{c}$} & \multicolumn{2}{c}{8} \\ \hline
\multicolumn{2}{l|}{$P_\mathrm{comp} (=2P_\mathrm{r}P_\mathrm{c})$} & \multicolumn{2}{c}{20,480} \\ \hline
\multicolumn{2}{l|}{$\lambda_\mathrm{comp}$ (cycles)} & \multicolumn{2}{c}{80} \\ \hline
\multicolumn{2}{l|}{$F_\mathrm{kernel}$(MHz)}& \multicolumn{2}{c}{281} \\ \hline
\multicolumn{2}{l|}{Comm. throughput (Gbps)}& \multicolumn{2}{c}{36} \\ \hline \hline
\multicolumn{2}{l|}{Cluster ID} & \#11 & \#12 \\ \hline
\multicolumn{2}{l|}{$N$} & 10,240 & 101,120 \\ \hline
\multicolumn{2}{l|}{$P_\mathrm{chip}$} & 8 & 79 \\ \hline
\multicolumn{2}{l|}{$N_\mathrm{hop} (=\lceil (P_\mathrm{chip}-1)/2 \rceil)$} & 4 & 39 \\ \hline
& $M_\mathrm{step}$ (cycles) & 872 & -- \\ \cline{2-4}
Measured & Performance (GMAC/s) & 33,787 & -- \\ \cline{2-4}
& $T_\mathrm{step}$ ($\mathrm{\mu}$s) & 3.1 & -- \\ \cline{2-4}
& Efficiency (\%) & 73.4 & -- \\ \hline
& $M_\mathrm{step}$ (cycles) & 856 & 7,026 \\ \cline{2-4}
Theory & Performance (GMAC/s) & 34,418 & 408,948 \\ \cline{2-4}
(Syn.) & $T_\mathrm{step}$ ($\mathrm{\mu}$s) & 3.0 & 25.0 \\ \cline{2-4}
& Efficiency (\%) & 74.8 & 90.0 \\ \hline
\end{tabular}
}
\end{table}

Fig. \ref{fig:cim}(b) shows the MAX-CUT score as a function of the computation time for the proposed architecture, along with those for the CIM and the simulated annealing (SA) from Ref. \cite{cim10}. The dashed line shows the MAX-CUT score (10,759,955) obtained by a greedy method (Sahni-Gonzales algorithm) as the baseline (target) score \cite{cim10}.
Similarly to \cite{cim10}, we solved the MAX-CUT problem (see Appendix A) 500 times with different initial values, and then selected the best one that reaches the baseline score in the shortest time. The time to reach the baseline score has been 475 $\mu$s (19 SB steps) for the proposed cluster, which is 1.2 times faster than that for the CIM (593 $\mu$s, 24 circulations).
Note that, in all 500 trials, the proposed architecture reached the baseline score within 21 SB steps.
Since the computation times per step (circulation) of the proposed architecture and the CIM are almost the same, the difference comes from the numbers of time-evolution steps to reach the baseline score.

\section{Related Works}
\label{sec:comp}
Several scalable Ising machines have been proposed, including one in our previous work \cite{multi_v1}. We describe the related works of multi-chip Ising machines (excluding single-chip ones and non-circuital ones).

One of the major classifications is on spin-to-spin connectivity; sparse coupled or fully coupled.
Most sparsely coupled Ising machines have regular and fixed local connections and thus they do not need all-to-all communication between chips.
However, they require mapping the original problems onto the fixed connections of the machines (minor-embedding).
Yamamoto \textit{et~al.} proposed a 1.3M-bit annealing system with 81 CMOS chips \cite{9634769}.
Its spin-to-spin connectivity is limited to a fixed sparse coupling called King's graph.
In contrast, our architecture is with full spin-to-spin connectivity, which does not require minor-embedding.
Another classification is digital computing versus analog computing.
Analog computing uses physical phenomena for computation.
Sharma \textit{et~al.} proposed a multi-chip physical Ising machine composed of nodal capacitors and coupling resistors \cite{ISCA22}.
They combined analog computing chips with digital inter-chip communication interfaces.
Their simulation showed 2,200 times faster results than our previous work \cite{multi_v1}.
They have not detailed assumed communication technology in their paper, but we assume that there should be a shorter chip-to-chip communication latency than that of our implementation.
In addition, as mentioned in the previous section, Honjo \textit{et~al.} proposed the 100,000-spin CIM, which is a state-of-the-art Ising machine \cite{cim10}.
They used optical phenomena supported by a matrix computing system on 56 FPGAs.
Our proposed architecture uses fully digital computing, which gives stable results.

Finally, we consider a fully digital Ising machine with full spin-to-spin connectivity, which is the same classification as our machine.
Muthumala and Hariyama proposed a multi-FPGA implementation of simulated quantum annealing (SQA) \cite{8918417}.
They used multiple FPGAs to extend the trotter size of SQA.
Their work showed a 290-fold speed improvement with two FPGAs compared with a single-core CPU implementation.
Yamamoto and Kawahara proposed a multi-chip annealing system with full spin-to-spin connectivity using pseudo-annealing \cite{YAMAMOTO}.
Their system can compute a 384-spin Ising problem with 17 Cyclone10 LP FPGAs at 10 MHz.
Kawamura \textit{et~al.} proposed RPA (ratio-controlled parallel annealing) \cite{amorphica}, which can solve the convergence issue of stochastic cellular automata annealing (SCA) \cite{STATICA} by introducing a ratio that controls the number of spins that can be flipped.
They also proposed and demonstrated a metamorphic annealing processor, called Amorphica\cite{amorphica}.
It can change the annealing algorithm to SA, DA, SCA, or RPA while solving problems.
It was implemented on ASICs and supported 2K-spin Ising problems.

\section{Conclusion}
\label{sec:conclusion}

We have demonstrated a scalable streaming architecture for a multi-chip implementation of SBMs with full spin-to-spin connectivity that allows enhancement of the computational performance up to an ideal upper limit determined only by the latency of chip-to-chip communication.

In the architecture, each chip corresponds to a partitioned subsystem of an oscillator (spin variable in SB) network and computes the time-evolved state of the subsystem while communicating with the other chips to share all the oscillator information at each SB time-evolution step. The multiple chips are connected in a dual-ring topology and the oscillator information is shared via a multi-hop communication process. Each chip calculates and accumulates the oscillator-to-oscillator interactions for the subsystem while receiving and sending the oscillator information (overlapping execution of computation and communication). The order and throughput of the processed data in chips are matched to those of transferred data between chips, and the transferred data are used only once (stream processing). The procedure of the SB time-evolution steps in each chip starts from calculating the interactions inside the self-subsystem and thus does not have to wait for the arrival of data from neighbors (hiding the initial chip-to-chip latency).

A performance model was constructed by quantitatively analyzing the computation cycles for the architecture and was validated on the series of experimental clusters. There are three operational modes depending on the architecture parameters. The mode changes from a computational throughput (determined by the computational parallelism) limiting mode to a communication latency limiting mode via an intermediate mode as the number of chips increases for a given problem size. In contrast, there is no mode characterized by the communication throughput (or the data transfer throughput) because the communication time is hidden by the overlapping execution of computation and communication. More precisely, the modes are identified by a relationship between the chip-to-chip communication latency (a fixed parameter for a given technology) and the characteristic time depending on the number of oscillators per chip and the computational parallelism per chip. The computational performance of the cluster increases linearly as the number of chips increases for a given problem size (ideal strong scaling property, at the computational throughput limiting mode) until the characteristic time decreases to be comparable to the chip-to-chip communication latency. The performance peaks at the intermediate mode and then begins to decline at the communication latency limiting mode. It is inevitable that the cluster performance is ultimately limited by the physical chip-to-chip latency.

Systematic experiments showed linear strong scaling of the performance up to the vicinity of the ideal communication latency limit determined only by the chip-to-chip communication latency. On a 16,384-node MAX-CUT problem, the performance reached 69,650 GMAC/s. This was 6.41-fold better than the performance of our previous architecture where the computation and communication were exclusive in the time domain \cite{multi_v1}. On a 32,768-node MAX-CUT problem, the 8-FPGA cluster showed a highest pipeline efficiency of 97.9\% and a highest measured performance of 77,775 GMAC/s. Furthermore, the performance of the 79-FPGA cluster on a 100,000-spin problem, projected by the theoretical model, is comparable to that of a state-of-the-art 100,000-spin optical Ising machine.

The cluster performance would be further enhanced by adopting advanced algorithms and hardware. In this work, adiabatic SB (aSB) was adopted to directly compare the proposed and previous architectures. Newer algorithms such as ballistic SB (bSB), discrete SB (dSB) \cite{sbm2}, heated ballistic SB (HbSB), and heated discrete SB (HdSB) \cite{Heated_SB} would improve performance. Larger on-chip memory, newer low-latency communication technology, and a reduced number of hops by changing the network topology would lead to performance improvement.

\appendix
\vspace{1cm}
\noindent \textbf{\Large Appendix}
\section{Ising problem}
The energy of the Ising model without external forces is defined by Eq. (\ref{eq:ising}).

\begin{align}
E(\boldsymbol{s})=-\cfrac{1}{2}\sum^N_{i=1}\sum^N_{i=1} J_{ij} s_i s_j \label{eq:ising}
\\ \text{where} \ s_i \in \{+1,-1\} \nonumber
\end{align}

The Ising problem is to find a combination of spins that minimizes the energy.
It is mathematically the same as the MAX-CUT problem \cite{FPL19}.
In our experiments, we used only the MAX-CUT problem whose weights were represented by 1 bit ($\pm 1$).
There have been no zero values except for the diagonal of the coefficient matrix (fully coupled).
Problems with sizes ranging from 1,024 to 32,768 were generated in the same way as in our previous work \cite{multi_v1}.
The 100,000-node problem was generated by the following command using rudy as in \cite{cim10}.
\\
\noindent {\centering \$ rudy -clique 100000 -random 0 1 55555 -times 2 -plus -1}\\

\section{Simulated bifurcation}
\label{sec:sb}
We describe the algorithm of SBM for single-chip implementations and multi-chip implementations.

The SBM internally represents the spins $\mathbf{s}$ of Ising models as oscillators (positions $\mathbf{x}$ and momenta $\mathbf{p}$).
In the simulation, the oscillators move based on the interaction of oscillators.
This time evolution is simulated in SB steps of time $\Delta t$, and the spins $\mathbf{s}$ are obtained by binarizing the final positions of the oscillators $\mathbf{x}$.
Furthermore, the simulation step is divided into $M$ sub-steps of time $\delta t$. 

The simulation is performed as shown in Algorithm \ref{algo:sb}, which is proposed in \cite{FPL19}.
The definitions of FX, FP, and $\Delta\alpha$ are as follows:
\begin{align}
\mathrm{FX}(x_i, h_i) &= \delta t \{ -(\alpha_0-\alpha)x_i - \beta_0x_i^3 - \eta h_i \}, \\
\mathrm{FP}(p_i) &= \delta t p_i,\\
\Delta \alpha &= (\alpha(\Delta t N_\mathrm{SBstep} ) - \alpha(0))/N_\mathrm{SBstep},
\end{align}
where $\alpha_0$, $\beta_0$, $\eta(t)$, and $\gamma_0$ are the parameters of the SB algorithms.

\begin{algorithm}[t]
\caption{Simulated bifurcation (SB)}
\begin{algorithmic}[1]
\STATE $\boldsymbol{x} \leftarrow \boldsymbol{0}$, $\alpha \leftarrow 0$
\STATE $\boldsymbol{p} \leftarrow $random\_vector($-0.1$ to $+0.1$)
\FOR {$l=1$ to $N_\mathrm{SBstep}$ }
\STATE //Matrix-vector multiplication (MM) \\
\STATE $\boldsymbol{\Delta p} = \boldsymbol{0}$
\FOR {$j=1$ to $N$}
\FOR {$i=1$ to $N$}
\STATE $\Delta p_i \leftarrow \Delta p_i + \Delta t \gamma_{0} J_{ij} x_j$ \\
\ENDFOR \\
\ENDFOR \\
\STATE //Time Evolution (TE) \\
\FOR {$i=1$ to $N$}
\STATE $p_{i} \leftarrow \Delta p_{i} + p_{i}$ \label{code:update_from}
\FOR {$m=1$ to $M$}
\STATE $p_{i} \leftarrow p_{i} + \mathrm{FX}(x_{i}, h_{i})$ \\
\STATE $x_{i} \leftarrow x_{i} + \mathrm{FP}(p_{i})$
\ENDFOR \\ \label{code:update_to}
\ENDFOR \\
\STATE $\alpha \leftarrow \alpha + \Delta \alpha$ \\
\ENDFOR
\end{algorithmic}
\label{algo:sb}
\end{algorithm}

We transformed Algorithm \ref{algo:sb} into Algorithm \ref{algo:part} for the implementation on clusters.
In Algorithm \ref{algo:part}, the order of MM and TE was swapped, and loop fusion was applied to loops of MM and TE.
The position data are shared between chips by communication (send and receive in Algorithm \ref{algo:part}).
Algorithm \ref{algo:part} shows pseudocode for single-ring clusters and does not represent data parallelism.
The "Update" function in the \ref{code:update}-th line represents the procedure of \ref{code:update_from} -- \ref{code:update_to} lines in Algorithm \ref{algo:sb}.
It is important to note that not all the input positions $\mathbf{x}$ need to be stored, but all the $\mathbf{\Delta p}$ needs to be stored.
Unlike the previous implementations \cite{FPL19, multi_v1}, the matrix-vector multiplication in this Algorithm \ref{algo:part} is processed in column-major order (prioritizing the processing of all the MAC operations within the same column before processing the MAC operations within the same row).
The for-loops in \ref{code:iter_j}-th line and \ref{code:iter_i}-th line are parallelizable.
The parallelism in \ref{code:iter_j}-th line is used for inter-chip parallelization.
The parallelism in \ref{code:iter_i}-th line is used for on-chip parallelization.

\begin{algorithm}[t]
\caption{Partitioned SB}
\begin{algorithmic}[1]
\STATE $\boldsymbol{x} \leftarrow \boldsymbol{0}$, $\boldsymbol{\Delta p} \leftarrow \boldsymbol{0}$, $\alpha \leftarrow 0$
\STATE $\boldsymbol{p} \leftarrow $random\_vector($-0.1$ to $+0.1$)
\FOR {$l=1$ to $N_\mathrm{SBstep}$ }
\STATE $\boldsymbol{\Delta p'} \leftarrow \boldsymbol{\Delta p}$
\STATE $\boldsymbol{\Delta p} \leftarrow \boldsymbol{0}$
\FOR {$j=1$ to $N$} \label{code:iter_j}
\STATE //Time Evolution (TE) and Receive operations \\
\IF {$j \le N/P_\mathrm{chip}$}
\STATE $x_{j}, p_{j} \leftarrow \mathrm{Update}(x_{j}, p_{j}, \Delta p'_{j}, h_{j})$ \label{code:update}
\STATE $x_\mathrm{tmp} \leftarrow x_{j}$
\ELSE
\STATE $x_\mathrm{tmp} \leftarrow \mathrm{receive}()$
\ENDIF \\
\STATE //Send operations
\IF {$j \le N - N/P_\mathrm{chip}$}
\STATE send($x_\mathrm{tmp}$)
\ENDIF \\
\STATE //Matrix-vector multiplication (MM) \\
\FOR {$i=1$ to $N/P_\mathrm{chip}$} \label{code:iter_i}
\STATE $\Delta p_{i} \leftarrow \Delta p_{i} + \Delta t \gamma_{0} J_{ij} x_\mathrm{tmp}$ \\
\ENDFOR \\
\ENDFOR \\
\STATE $\alpha \leftarrow \alpha + \Delta \alpha$ \\
\ENDFOR
\end{algorithmic}
\label{algo:part}
\end{algorithm}

\section{Implementation}
\begin{figure}[t]
\includegraphics[width=\linewidth]{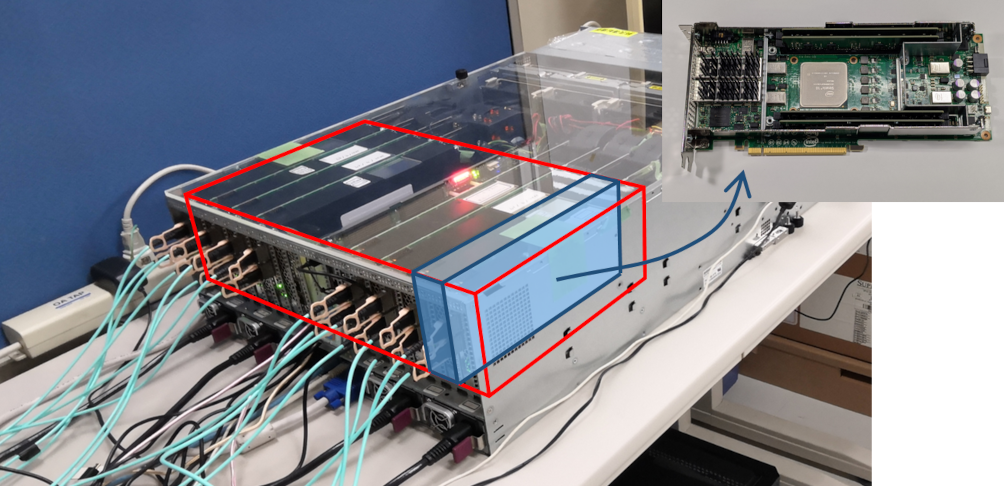}
\caption{
An eight-chip cluster on a 4U-sized server\label{fig:pic}
}
\end{figure}

We implemented the proposed architecture on Stratix10 FPGAs.
Fig. \ref{fig:pic} shows the cluster with eight FPGAs installed in a 4U-sized server.
The model of the FPGA boards is Intel FPGA Programmable Acceleration Card D5005 \cite{D5005} with 2800K-LE Stratix 10 FPGA.
Each board has two QSFP28 ports that are capable of 100 Gbps communication.
We used high speed peer-to-peer serial link network for communication \cite{SLIII}.
For more details, see Section S1 and S2 in Supplementary Material.

We used a High-Level Synthesis (HLS) tool \cite{openclsdk_d5005} for generating Register-Transfer Level (RTL) codes.
The synthesized architecture parameters ($\lambda_\mathrm{TE}$, $\lambda_\mathrm{MM}$, and $\lambda_\mathrm{TX}$) were obtained from HLS reports and estimation of the latency of inter-module FIFO.
$\lambda_\mathrm{PHY}$ (340 ns) was obtained by fitting the theoretical model cycles to the measured cycles of experiments using S1K designs (\#1, \#2, and \#3).

In the implementations, each coupling coefficient $J_{ij}$ was coded in one bit, and each position $x_{i}$ and momenta $p_{i}$ were coded in 16 bits (similarly to \cite{FPL19, multi_v1}).
The implemented machine can calculate problems of smaller sizes than supported.
The problem size can be set before the calculation.
The execution time does not change even if a smaller problem is used.

\section*{Acknowledgment}
The authors thank Hayato Goto and Yohei Hamakawa for their valuable comments and support.

\section*{Competing interests}
T.K., M.Y., R.H., and K.T. are included in inventors on two U.S. patent applications related to this work filed by the Toshiba Corporation (no. 16/118646, filed 31 August 2018; no. 17/680213, filed 24 February 2022). The authors declare that they have no other competing interests.

\begin{figure}[H]
\vspace{0.5cm}
\noindent\includegraphics[width=1in,height=1.25in,clip,keepaspectratio]{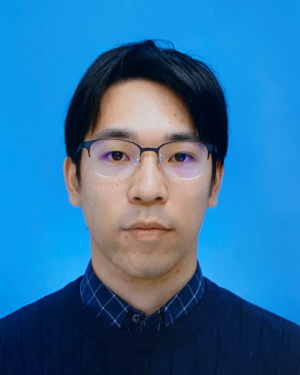}\\
{\small Tomoya Kashimata received the B.E., and M.E. degrees in computer science and engineering from Waseda University, Japan, in 2018 and 2020, respectively. He joined the Corporate Research and Development Center, Toshiba Corporation, Japan, in 2020. His research interests include computer architecture, reconfigurable architecture, and processing-in-memory.}
\vspace{-0.5cm} 
\end{figure}

\begin{figure}[H]
\vspace{0.5cm}
\noindent\includegraphics[width=1in,height=1.25in,clip,keepaspectratio]{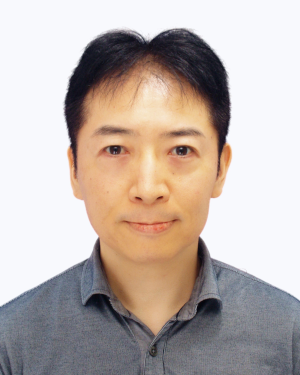}\\
{\small Masaya Yamasaki received the B.E. and M.E. degrees in computer science and communication engineering from Kyushu University, Japan, in 1997 and 1999, respectively. He joined Toshiba Corporation, in 1999. He was engaged in the development of image processing engines (interframe interpolation technology) for digital televisions (including ones with Cell Broadband Engines) and FPGA-based coprocessors for multi-channel video recording, three-dimensional display, and industrial systems. His research interests include domain-specific computing, high-level synthesis design space exploration, and proof-of-concept study with FPGA devices.}
\vspace{-0.5cm} 
\end{figure}

\begin{figure}[H]
\vspace{0.5cm}
\noindent\includegraphics[width=1in,height=1.25in,clip,keepaspectratio]{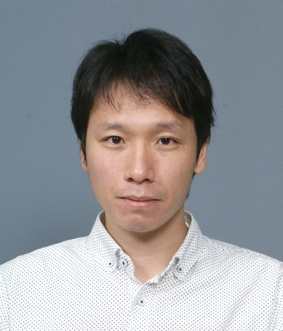}\\
{\small Ryo Hidaka received the B.E. and M.E. degrees in systems design and informatics from the Kyushu Institute of Technology, Japan, in 2006 and 2008, respectively. He joined Toshiba Corporation, in 2008. He was engaged in the development of main processors (2D-to-3D conversion and local dimming) for digital televisions, an image recognition processor called Visconti\textsuperscript{TM}, and host controllers for flash-memory cards. His current research interests include domain-specific computing, high-level synthesis design methodology, and proof-of-concept study with FPGA devices.}
\vspace{-0.5cm} 
\end{figure}

\begin{figure}[H]
\vspace{0.5cm}
\noindent\includegraphics[width=1in,height=1.25in,clip,keepaspectratio]{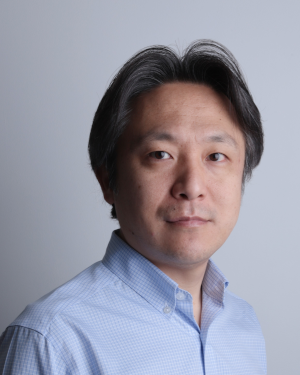}\\
{\small Kosuke Tatsumura received the B.E., M.E., and Ph.D. degrees in electronics, information and communications engineering from Waseda University, Japan, in 2000, 2001, and 2004, respectively. After working as a Postdoctoral Fellow with Waseda University, he joined Toshiba Corporation, in 2006. He is a Chief Research Scientist, leading a research team and several projects toward realizing innovative industrial systems based on cutting-edge computing technology. He was a member of the Emerging Research Devices (ERD) Committee in the International Technology Roadmap for Semiconductors (ITRS), from 2013 to 2015. He has been a Lecturer with Waseda University, since 2013. He was a Visiting Researcher with the University of Toronto, from 2015 to 2016. He received the Best Paper Award at IEEE International Conference on Field-Programmable Technology (FTP), in 2016. His research interests include domain-specific computing, quantum/quantum-inspired computing, and their applications.}
\vspace{-0.5cm} 
\end{figure}

\pagebreak

\setcounter{section}{0}
\setcounter{table}{0}
\setcounter{figure}{0}
\setcounter{equation}{0}

\newcolumntype{Y}{>{\centering\arraybackslash}X}
\renewcommand{\theequation}{S\arabic{equation}}
\renewcommand{\thefigure}{S\arabic{figure}}
\renewcommand{\thetable}{S\arabic{table}}
\renewcommand{\thesection}{S\arabic{section}}

\newcolumntype{C}[1]{>{\centering\arraybackslash}p{#1}}
\newcolumntype{L}[1]{>{\raggedright\arraybackslash}p{#1}}
\newcolumntype{R}[1]{>{\raggedleft\arraybackslash}p{#1}}

\twocolumn[
\textbf{\LARGE Supplementary Material of\\
Efficient and Scalable Architecture for Multiple-chip Implementation of Simulated Bifurcation Machines}\\
{\centering\small Tomoya Kashimata, Masaya Yamasaki, Ryo Hidaka, and Kosuke Tatsumura$^{\ast}$\\}
{\centering\scriptsize  Corporate Research and Development Center, Toshiba Corporation, \\
 1 Komukai Toshiba-cho, Saiwai-ku, Kawasaki 212-8582, Japan\\}
\vspace{0.8cm}]

\section{Implementation on Arria10 FPGAs}
We implemented the proposed clusters on Intel Arria10 FPGAs (Aval APX-AA10L1 \cite{Saval}).
Each FPGA board has an Arria10 GX 1150 FPGA \cite{SArria10} and two QSFP+ connectors.
For the implementation, we used an OpenCL high-level synthesis tool (Intel FPGA SDK for OpenCL Pro Edition 17.1 \cite{Sopenclsdk_v1}).
Table \ref{tab:a10_result} shows the measured and theoretical performance and the implementation details including the resource usage.
The physical environment (CPU-server, FPGA boards and communication modules) and the number of MAC units per chip are the same as in our previous work \cite{Smulti_v1}.
The performance for a 16,384-spin problem in this work reached 17,959 GMAC/s (Cluster ID: \#A9), which is 1.65 times higher than the corresponding one in our previous work \cite{Smulti_v1}.
In addition, the peak performance for an 8,192-spin problem reached 17,451 GMAC/s (Cluster ID: \#A6), which is 2.63 times higher than the corresponding one in our previous work \cite{Smulti_v1}.
We determined the typical parameters as $\lambda_\mathrm{comm} = 161$ and $\lambda_\mathrm{comp} = 64$ from the module latency reported by HLS, the measured internal FIFO latency, and the measured chip-to-chip latency.
For the comparison with our previous work, we also assumed that the clock frequency ($F_\mathrm{kernel}$) for the typical parameter is 266 MHz.
The on-chip RAM usage is determined by the implemented matrix size.
The number of rows of the matrix corresponds to the number of oscillators for which the chip is responsible ($N/P_\mathrm{chip}$), and the number of columns indicates the maximum number of oscillators ($N$) that the clusters can compute.

\begin{table*}
\caption{Measured and theoretical computational performance and implementation details on the proposed Arria10 clusters\label{tab:a10_result}}
\centering
\resizebox{\linewidth}{!}{
\begin{tabular} {l|L{1.8cm}|ccc|ccc|ccc|C{1.3cm}|C{1.2cm}}
\hline
\multicolumn{2}{l|}{Design ID                   } & \multicolumn{3}{c|}{A512                   }     & \multicolumn{3}{c|}{A1K                   }  & \multicolumn{3}{c|}{A2K                     }  & Typical\par parameter& Board\par capacity  \\ \hline
\multicolumn{2}{l|}{$N/P_{\rm chip} (=P_{\rm r})$}& \multicolumn{3}{c|}{512                    }     & \multicolumn{3}{c|}{1,024                 }  & \multicolumn{3}{c|}{2,048                   }  & -                & - \\ \hline
\multicolumn{2}{l|}{Operational mode            } & \multicolumn{3}{c|}{Communication latency  }   & \multicolumn{3}{c|}{Intermediate mode}       & \multicolumn{3}{c|}{Computational throughput}  & -                & - \\
\multicolumn{2}{l|}{of the model                } & \multicolumn{3}{c|}{ limiting mode }             & \multicolumn{3}{c|}{                      }  & \multicolumn{3}{c|}{limiting mode           }  &                  &  \\ \hline
\multicolumn{2}{l|}{$M_{\rm compelem}$          } & \multicolumn{3}{c|}{32                     }     & \multicolumn{3}{c|}{128                   }  & \multicolumn{3}{c|}{512                     }  & -                & - \\ \hline
\multicolumn{2}{l|}{$\lambda_{\rm comm}$        } & \multicolumn{3}{c|}{157                    }     & \multicolumn{3}{c|}{165                   }  & \multicolumn{3}{c|}{160                     }  & 161              & - \\ \hline\hline
\multicolumn{2}{l|}{$P_{\rm c}$                 } & \multicolumn{3}{c|}{8                      }     & \multicolumn{3}{c|}{4                     }  & \multicolumn{3}{c|}{2                       }  & -                & - \\ \hline
\multicolumn{2}{l|}{$P_{\rm comp} (=2P_{\rm r}P_{\rm c})$} & \multicolumn{9}{c|}{8,192  }                                                                                                            & 8,192            & - \\ \hline
\multicolumn{2}{l|}{$\lambda_{\rm comp}$        } & \multicolumn{3}{c|}{64                     }     & \multicolumn{3}{c|}{64                    }  & \multicolumn{3}{c|}{64                      }  & 64               & - \\ \hline
\multicolumn{2}{l|}{$\lambda_{\rm TE}$          } & \multicolumn{3}{c|}{45                     }     & \multicolumn{3}{c|}{46                    }  & \multicolumn{3}{c|}{47                      }  & -                & - \\ \hline
\multicolumn{2}{l|}{$\lambda_{\rm JX}$          } & \multicolumn{3}{c|}{19                     }     & \multicolumn{3}{c|}{18                    }  & \multicolumn{3}{c|}{17                      }  & -                & - \\ \hline
\multicolumn{2}{l|}{$\lambda_{\rm TX}$          } & \multicolumn{3}{c|}{8                      }     & \multicolumn{3}{c|}{8                     }  & \multicolumn{3}{c|}{8                       }  & -                & - \\ \hline
\multicolumn{2}{l|}{$\lambda_{\rm PHY}$ (ns)    } & \multicolumn{9}{c|}{375                    }                                                                                                     & -                & - \\ \hline
\multicolumn{2}{l|}{$\lambda_{\rm PHY}$ (cycle) } & \multicolumn{3}{c|}{104                    }     & \multicolumn{3}{c|}{111                   }  & \multicolumn{3}{c|}{105                     }  & -                & - \\ \hline
\multicolumn{2}{l|}{$F_{\rm kernel}$(MHz)       } & \multicolumn{3}{c|}{275.0                  }     & \multicolumn{3}{c|}{294.4                 }  & \multicolumn{3}{c|}{278.6                   }  & 266.0            & - \\ \hline
\multicolumn{2}{L{2.5cm}|}{Communication Throughput (Gbps)} & \multicolumn{3}{c|}{35                  }     & \multicolumn{3}{c|}{19                    }  & \multicolumn{3}{c|}{9                       }  & -                & - \\ \hline \hline
\multicolumn{2}{l|}{ALUTs                       } & \multicolumn{3}{c|}{318,218                }     & \multicolumn{3}{c|}{ 320,755              }  & \multicolumn{3}{c|}{ 334,858                }  & -                & - \\ \hline
\multicolumn{2}{l|}{Logic utilization           } & \multicolumn{3}{c|}{ 190,215 ( 45 \% )     }     & \multicolumn{3}{c|}{ 191,705 ( 45 \% )    }  & \multicolumn{3}{c|}{ 200,484 ( 47 \% )      }  & -                & 427,200 \\ \hline
\multicolumn{2}{l|}{DSP blocks                  } & \multicolumn{3}{c|}{ 272 ( 18 \% )         }     & \multicolumn{3}{c|}{ 136 ( 9 \% )         }  & \multicolumn{3}{c|}{ 68 ( 4 \% )            }  & -                & 1,518 \\ \hline
\multicolumn{2}{l|}{Memory bits                 } & \multicolumn{3}{c|}{ 13,762,536 ( 25 \% )  }     & \multicolumn{3}{c|}{ 22,163,048 ( 40 \% ) }  & \multicolumn{3}{c|}{ 38,978,152 ( 70 \% )   }  & -                & 55,562,240 \\ \hline
\multicolumn{2}{l|}{RAM blocks                  } & \multicolumn{3}{c|}{ 1,037 ( 38 \% )       }     & \multicolumn{3}{c|}{ 1,521 ( 56 \% )      }  & \multicolumn{3}{c|}{ 2,529 ( 93 \% )        }  & -                & 2,713 \\ \hline
\multicolumn{2}{l|}{Submatrix $J$ size per chip } & \multicolumn{3}{c|}{$512 \times 16,384$    }     & \multicolumn{3}{c|}{$1,024 \times 16,384$ }  & \multicolumn{3}{c|}{$2,048 \times 16,384$   }  & -                & - \\ \hline\hline
\multicolumn{2}{l|}{Cluster ID                  } & \#A1                 &  \#A2   & \#A3            & \#A4                  & \#A5   & \#A6        & \#A7                  & \#A8         & \#A9    & -                & - \\ \hline
\multicolumn{2}{l|}{$N$                         } & 1,024                &  2,048  & 4,096           & 2,048                 & 4,096  & 8,192       & 4,096                 & 8,192        & 16,384  & -                & - \\ \hline
\multicolumn{2}{l|}{$P_{\rm chip}$              } & 2                    &  4      & 8               & 2                     & 4      & 8           & 2                     & 4            & 8       & -                & - \\ \hline
\multicolumn{2}{l|}{$N_{\rm hop}$               } & 1                    &  2      & 4               & 1                     & 2      & 4           & 1                     & 2            & 4       & -                & - \\ \hline
                       &  $M_{\rm step}$ (cycles) & 250                  &  404    & 714             & 361                   & 619    & 1,132       & 1,088                 & 2,112        & 4,164   & -                & - \\ \cline{2-13}
Measured               & Performance (GMAC/s)     & 1,152                &  2,854  & 6,460           & 3,419                 & 7,977  & 17,451      & 4,295                 & 8,851        & 17,959  & -                & - \\ \cline{2-13}
                       &  $T_{\rm step}$ ($\mu$s) & 0.9                  &  1.5    & 2.6             & 1.2                   & 2.1    & 3.8         & 3.9                   & 7.6          & 14.9    & -                & - \\ \cline{2-13}
                       &  Efficiency (\%)         & 25.6\%               &  31.7\% & 35.9\%          & 70.9\%                & 82.7\% & 90.5\%      & 94.1\%                & 97.0\%       & 98.4\%  & -                & - \\ \hline
Theory                 &  $M_{\rm step}$          & 253                  &  410    & 724             & 357                   & 613    & 1,125       & 1,088                 & 2,112        & 4,160   & -                & - \\ \cline{2-13}
(Syn.)                 & Performance (GMAC/s)     & 1,139                &  2,812  & 6,371           & 3,457                 & 8,055  & 17,560      & 4,295                 & 8,851        & 17,976  & -                & - \\ \hline
Theory                 &  $M_{\rm step}$          & 257                  &  418    & 740             & 353                   & 609    & 1,121       & 1,088                 & 2,112        & 4,160   & -                & - \\ \cline{2-13}
(Typ.)                 &  GMAC/s                  & 1,084                &  2,668  & 6,029           & 3,159                 & 7,326  & 15,922      & 4,101                 & 8,451        & 17,163  & -                & - \\ \hline
\end{tabular}
}
\end{table*}

\section{Implementation on Stratix10 FPGAs}

Table \ref{tab:s10_resource} shows the resource usage and detailed latency of the synthesized designs for the Stratix10 FPGAs.
The logic utilization increases with increasing $N/P_\mathrm{chip}$, even though the numbers of MAC units are the same.
Larger $N/P_\mathrm{chip}$ requires more accumulators, which can increase the logic utilization.
The designs consume few DSPs because they are not required for the matrix vector multiplication.
For S4K, S8K, and S1280 designs, the  $J$ submatrix sizes were determined by the memory capacity.
For S1K and S2K designs, the $J$ submatrix sizes were determined based on our assumption that $P_\mathrm{chip}$ is at most 16.
As for the RAM block usage, S1K and S2K designs still have room to expand the $J$ submatrix sizes more than 2-fold.

We reduced the number of MAC units in S1280 design (to be 20,480 MAC units) compared with the other designs (32,768 MAC units).
Both before and after the reduction, S1280 designs have been at Mode C (the communication latency limiting mode).
At Mode C, the MAC units can be reduced without significant performance degradation.

\begin{table*}
\caption{Implementation details on the proposed Stratix10 clusters \label{tab:s10_resource}}
\centering
\resizebox{\linewidth}{!}{
\begin{tabular} {l|c|c|c|c|c|c}
\hline
Design ID	& S1K	& S2K	& S4K	& S8K	& S1280	& Board capacity \\ \hline
$N/P_{\rm chip}$ (=$P_{\rm r}$)	& 1,024	& 2,048	& 4,096	& 8,192	& 1,280	& -  \\ \hline
%Maximum N supported	& 16,384	& 32,768	& 32,768	& 16,384	& 102,400	& - \\ \hline
$P_{\rm comp}$ & \multicolumn{4}{c|}{32,768} & 20,480 & - \\ \hline
$\lambda_{\rm TE}$	& 61	& 58	& 58	& 58	& 58	& - \\ \hline
$\lambda_{\rm JX}$	& 20	& 22	& 29	& 43	& 22	& - \\ \hline
$\lambda_{\rm TX}$	& 20	& 20	& 15	& 15	& 20	& - \\ \hline
$\lambda_{\rm PHY}$ (ns)	& \multicolumn{5}{c|}{340}		& - \\ \hline
$\lambda_{\rm PHY}$ (cycle)	& 96	& 103	& 104	& 94	& 96	& - \\ \hline
$\lambda_{\rm comp}$	& 81	& 80	& 87	& 101	& 80	& - \\ \hline
$\lambda_{\rm comm}$	& 177	& 181	& 177	& 167	& 174	& - \\ \hline
$F_{\rm kernel}$(MHz)	& 281	& 301	& 303	& 275	& 281	& - \\ \hline
ALUTs	& 850,294	& 872,787	& 966,232	& 1,118,459	& 661,642	& -- \\ \hline
Logic utilization	& 592,910	& 604,821	& 652,402	& 728,726	& 484,884	& 933,120 \\ 
	&  ( 64 \% )	&  ( 65 \% )	&  ( 70 \% )	&  ( 78 \% )	&  ( 52 \% )	&  \\ \hline
DSP blocks	& 240	& 120	& 60	& 30	& 120	& 5,760 \\ 
	&  ( 4 \% )	&  ( 2 \% )	&  ( 1 \% )	&  ( $<$ 1 \% )	&  ( 2 \% )	&  \\ \hline
Memory bits	& 25,906,808	& 75,240,568	& 141,947,512	& 142,212,064	& 176,301,688	& 240,046,080 \\ 
	&  ( 11 \% )	&  ( 31 \% )	&  ( 59 \% )	&  ( 59 \% )	&  ( 73 \% )	&  \\ \hline
RAM blocks	& 2,007	& 4,977	& 8,993	& 9,101	& 11,163	& 11,721 \\ 
	&  ( 17 \% )	&  ( 42 \% )	&  ( 77 \% )	&  ( 78 \% )	&  ( 95 \% )	&  \\ \hline
Submatrix $J$ size per chip & $1,024\times 16,384$ & $2,048\times 32,768$ & $4,096 \times 32,768$ & $8,192 \times 16,384$ & $1,280 \times 102,400$ & - \\ \hline
\end{tabular}
}
\end{table*}

\section{Details of the measurements}
In our experiments, all positions ($\boldmath{x}$) were initialized to zero and all momenta ($\boldmath{p}$) were randomly initialized between -0.1 and +0.1. The algorithm used for generating the pseudorandom numbers was Mersenne Twister \cite{Smatsumoto98}.
In the experiment for comparison with the CIM, the seed of the pseudorandom number was set to 1, then it was repeated 500 times and the best result was employed.

$M_\mathrm{step}$ was measured based on the operation clock cycles of the TE module.
The clock cycle counts were obtained by a profiler tool contained in the OpenCL development environment.
To minimize the measurement variability, we measured the number of clock cycles on the FPGA chip that started last.
We also used the minimum number in 10 repeated measurements.
The measurement was also repeated by changing the number of SB steps ($N_\mathrm{step}$), and we obtained $M_\mathrm{step}$ from the slope of the relationship between the measured clock cycles and $N_\mathrm{step}$.

\section{Communication throughput bound}
The required communication throughput for this architecture was calculated from $P_\mathrm{c}$ and the clock frequency ($F_\mathrm{kernel}$).
In the implementation, a position datum is expressed by a 16-bit variable.
Hence, the required throughput was calculated as $16P_\mathrm{c}F_\mathrm{kernel}$.
Note that both $P_\mathrm{c}$ and $F_\mathrm{kernel}$ are determined by the circuit design and do not depend on the cluster size.
For example, in S1K design, the required throughput has been 71.9 Gbps ($P_\mathrm{c} = 16$ and $F_\mathrm{kernel} = 281 \mathrm{MHz}$), which has been the highest in this work.
Since the Stratix10 FPGAs have network interfaces capable of 100-Gbps throughput, the communication throughput does not limit the performance.

\section{Memory throughput bound}
The performance of the proposed clusters in this work is not limited by the memory throughput, but the total computational parallelism ($P_\mathrm{comp}$) is determined by considering the memory throughput since the computational modules and the on-chip memories are tightly coupled.
Note that when the memory throughput limits the parallelism, the performance changes in the same way that computational parallelism decreases. Furthermore, in this work, the memory latency is already included in the computational latency ($\lambda_\mathrm{comp}$).

\section{Details of the strong scaling characteristics}
\subsection{Intermediate mode}
We show that the performance increases as the number of chips increases in the intermediate mode.
$M_\mathrm{step}$ in the intermediate mode is expressed as
\begin{align}
M_{\rm step} = (P_{\rm chip}-1) M_{\rm compelem} + \lambda_{\rm comm} + \lambda_{\rm comp}.
\end{align}
From the definition of $M_\mathrm{compelem}$, the following equation is derived: 
\begin{align}
M_{\rm step} = \cfrac{N}{2P_{\rm c}}\left(1-\cfrac{1}{P_{\rm chip}}\right) + \lambda_{\rm comm} + \lambda_{\rm comp}.
\end{align}
Since $P_\mathrm{c}$ is proportional to $P_\mathrm{chip}$ when $P_\mathrm{comp}$ is fixed, we can transform it using $P_\mathrm{c} = kP_\mathrm{chip}$ with an appropriate positive coefficient $k$.
\begin{align}
M_{\rm step} = \cfrac{N}{2k}\left(\cfrac{1}{P_{\rm chip}}-\cfrac{1}{P_{\rm chip}^{2}}\right) + \lambda_{\rm comm} + \lambda_{\rm comp}
\end{align}
Differentiating this equation by $P_\mathrm{chip}$ gives the following equation.
\begin{align}
\cfrac{\mathrm{\partial}M_{\rm step}}{\mathrm{\partial}P_{\rm chip}} = \cfrac{N}{2k}\left(-\cfrac{1}{P_{\rm chip}^{2}}+\cfrac{2}{P_{\rm chip}^{3}}\right) \label{eq:this188}
\end{align}
The condition for Eq. \ref{eq:this188} to be less than zero (i.e., for the performance to monotonically increase with increasing $P_\mathrm{chip}$) is derived to be $P_\mathrm{chip} > 2$.
Because the use of multiple chips is assumed here, the performance increases with increasing $P_\mathrm{chip}$ under the intermediate mode.

\subsection{Communication latency limiting mode}
We show that the performance decreases as $P_\mathrm{chip}$ increases in the communication latency limiting mode. In this mode, $M_\mathrm{step}$ is expressed as
\begin{align}
M_{\rm step} = N_{\rm hop}\lambda_{\rm comm} + N_{\rm lastelem} M_{\rm compelem} + \lambda_{\rm comp}.
\end{align}
Now, we consider only the case where $P_\mathrm{chip}$ is an even number ($N_\mathrm{lastelem} = 1$ and $N_\mathrm{hop} = P_\mathrm{chip}/2$ in this case).
\begin{align}
M_{\rm step} = \cfrac{P_{\rm chip}\lambda_{\rm comm}}{2} + \cfrac{N}{2P_{\rm chip}P_{\rm c}} + \lambda_{\rm comp}
\end{align}
The condition for the partial derivative with respect to $P_\mathrm{chip}$ to be positive (i.e., where the performance decreases as $P_\mathrm{chip}$ increases) is expressed as follows: 
\begin{align}
\cfrac{\mathrm{\partial}M_{\rm step}}{\mathrm{\partial}P_{\rm chip}} = \cfrac{\lambda_{\rm comm}}{2} - \cfrac{N}{2P_{\rm c}P_{\rm chip}^2} &> 0 \\
P_{\rm chip} &> \cfrac{2M_{\rm compelem}}{\lambda_{\rm comm}} \label{eq:cll_cond}
\end{align}
Because the condition, $2M_\mathrm{compelem} < \lambda_\mathrm{comm}$, is always satisfied in this mode, Eq. \ref{eq:cll_cond} always holds true.
Therefore, the performance decreases monotonically with increasing $P_\mathrm{chip}$.
When $P_\mathrm{chip}$ is an odd number, the same result is obtained by almost the same argument.

\section{Details of the ideal performance}
We show that the ideal performance for the computation is proportional to $P_\mathrm{chip}$, and the ideal performance for the communication is inversely proportional to $P_\mathrm{chip}$.
In this paper, the ideal performance has been defined as the performance determined solely by the computational parallelism or the chip-to-chip communication latency.

As described in the main text, the ideal number of clock cycles based on the computational throughput is $P_\mathrm{chip}M_\mathrm{compelem}$.
Assuming that the total number of MAC operations per SB step is $N^2$ for simplicity, the ideal performance, $Q_{\rm comp}^{\rm ideal}$, is calculated as follows.
\begin{align}
Q_{\rm comp}^{\rm ideal} &=& \cfrac{N^2}{P_{\rm chip} M_{\rm compelem}} F_{\rm kernel}\\
                         &=& P_{\rm comp} F_{\rm kernel} \cdot P_{\rm chip} \label{eq:prod}
\end{align}
Equation \ref{eq:prod} concisely shows that ideal performance is equal to the product of the total parallelism and the clock frequency.
Thus, the performance appears to be proportional to $P_\mathrm{chip}$.

The ideal number of cycles based on the communication latency is given by $N_\mathrm{hop}\lambda_\mathrm{comm}$.
The ideal performance, $Q_{\rm comm}^{\rm ideal}$, is expressed as  
\begin{align}
Q_{\rm comm}^{\rm ideal} = \cfrac{N^2}{N_{\rm hop} \lambda_{\rm comm}} F_{\rm kernel}.
\end{align}
Assume that the $P_\mathrm{chip}$ is an even number, then $N_\mathrm{hop}$ = $P_\mathrm{chip}$/2.
\begin{align}
Q_{\rm comm}^{\rm ideal} = \cfrac{2N^2}{\lambda_{\rm comm}}F_{\rm kernel}\cdot\cfrac{1}{P_{\rm chip}} 
\label{eq:ideal_comm}
\end{align}
Equation \ref{eq:ideal_comm} shows that the ideal performance based on the communication latency is inversely proportional to $P_\mathrm{chip}$.
When $P_\mathrm{chip}$ is an odd number,  $Q_{\rm comm}^{\rm ideal}$ shifts along $P_\mathrm{chip}$.

\end{document}